\documentclass[useAMS,usenatbib]{mn2e} \usepackage{natbib}
\usepackage{graphicx} \newcommand{\msun}{\mbox{$\,{\rm M}_\odot$}}

\title[Peculiarities in Velocity Dispersion and Surface Density
Profiles of Star Clusters]{Peculiarities in Velocity Dispersion and
Surface Density Profiles of Star Clusters}

\author[A.H.W. K\"upper, P. Kroupa, H. Baumgardt and
  D.C. Heggie]{Andreas H.W. K\"upper$^{1,2}$\thanks{E-mail:
  \mbox{akuepper@astro.uni-bonn.de} (AHWK);
  \mbox{pavel@astro.uni-bonn.de} (PK); \mbox{h.baumgardt@uq.edu.au}
  (HB); \mbox{d.c.heggie@ed.ac.uk} (DCH)}, Pavel Kroupa$^1$, Holger
  Baumgardt$^{1,3}$ and Douglas C. Heggie$^{4}$\\ $^{1}$Argelander
  Institut f\"ur Astronomie (AIfA), Auf dem H\"ugel 71, 53121 Bonn,
  Germany\\ $^{2}$European Southern Observatory, Alonso de Cordova
  3107, Vitacura, Santiago, Chile\\ $^{3}$University of Queensland,
  School of Mathematics and Physics, QLD 4072, Australia\\
  $^{4}$University of Edinburgh, School of Mathematics and Maxwell
  Institute for Mathematical Sciences, King's Buildings, Edinburgh EH9
  3JZ, UK}

\begin{document}

\date{Accepted \ldots. Received \ldots; in original form \ldots}

\pagerange{\pageref{firstpage}--\pageref{lastpage}} \pubyear{2009}

\maketitle

\label{firstpage}

\maketitle

\begin{abstract}
Based on our recent work on tidal tails of star clusters
\citep{Kuepper09} we investigate star clusters of a few $10^4 \msun$
by means of velocity dispersion profiles and surface density
profiles. We use a comprehensive set of $N$-body computations of star
clusters on various orbits within a realistic tidal field to study the
evolution of these profiles with time, and ongoing cluster
dissolution.\\ 
From the velocity dispersion profiles we find that the
population of potential escapers, i.e. energetically unbound stars
inside the Jacobi radius,  dominates clusters at radii above about
50\% of the Jacobi radius. Beyond 70\% of the Jacobi radius nearly all
stars are energetically unbound. The velocity dispersion therefore
significantly deviates from the predictions of simple equilibrium
models in this regime. We furthermore argue that for this reason this part of a cluster cannot be used to detect a dark matter halo or deviations from Newtonian gravity.\\ 
By fitting templates to the about $10^4$ computed
surface density profiles we estimate the accuracy which can be
achieved in reconstructing the Jacobi radius of a cluster in this
way. We find that the template of King (1962) works well for extended
clusters on nearly circular orbits, but shows significant flaws in
the case of eccentric cluster orbits. This we fix by extending this
template with 3 more free parameters. Our template can reconstruct the
tidal radius over all fitted ranges with an accuracy of about 10\%, and
is especially useful in the case of  cluster
data with a wide radial coverage  and for clusters showing significant extra-tidal stellar
populations. No other template that we have tried can yield comparable
results over this range of cluster conditions. All templates fail to
reconstruct tidal parameters of concentrated clusters, however.\\
Moreover, we find that the bulk of a cluster adjusts to the mean tidal
field which it experiences and not to the tidal field at
perigalacticon as has often been assumed in other investigations,
i.e. a fitted tidal radius is a cluster's time average mean tidal
radius and not its perigalactic one.\\ Furthermore, we study the tidal
debris in the vicinity of the clusters and find it to be well
represented by a power-law with a slope of -4 to -5. This steep slope
we ascribe to the epicyclic motion of escaped stars in the tidal
tails. Star clusters close to apogalacticon show a significantly
shallower slope of up to -1, however. We suggest that  clusters at apogalacticon
can be identified by measuring this
slope.
\end{abstract}


\section{Introduction}
Velocity dispersion profiles and surface density profiles are among
the most basic tools for investigating the structure of star
clusters. However, such investigations indicate that the
region around the tidal radius, at which the internal acceleration of a
star cluster is similar to the tidal acceleration due to the galactic
tidal field, is particularly poorly understood.\\

Velocity dispersion profiles sometimes show peculiarities which have been 
 discussed in the literature.  \citet{Drukier98}, for example, observed a flattening
in the outer parts of the velocity dispersion profile of the Galactic
globular cluster M15 which they interpreted as an effect of tidal
heating by the general Galactic tide or by tidal
shocks. \citet{Scarpa03} also found a flattening of the velocity
dispersion profile for $\omega$ Cen and more recently for other
Galactic globular clusters like NGC6171, NGC7099 and NGC288
\citep{Scarpa07}. The deviation from an expected Keplerian fall-off in
the velocity dispersion profile occurred in all clusters at radii
where the internal gravitational acceleration is about $a_0 = 1.2
\times 10^{-8}\mbox{cm s}^{-2}$ and was therefore interpreted by
Scarpa et al.~as a hint of Modified Newtonian Dynamics
\citep{Milgrom83} in globular clusters. Alternatively, they briefly
discussed the possible effect of tidal heating or a dark matter halo
on the cluster stars.\\\

On the contrary, \citet{McLaughlin03b} found that by fitting a
\citet{Wilson75} model, which has a less sharp cut-off at the tidal
radius than the commonly used \citet{King66} model, to the re-analyzed
$\omega$ Cen data, its velocity dispersion profile could be explained
without modifying Newtonian gravity and without adding dark
matter. Similar investigations of cluster profiles,
e.g. \citet{Lane09, Lane10} and \citet{Baumgardt10}, were also not in
favour of MOND.\\

But it is not only the velocity dispersion profiles of star clusters which behave
strangely at the tidal boundary;  their surface density profiles also
show peculiarities and sometimes controversial behaviour.\\

\citet{King62} showed that the surface density profiles of many
globular clusters can be fitted by a simple analytical formula having
a sharp cut-off radius, which could be interpreted as the tidal radius
of the cluster. Later he derived a set of physically motivated models,
with a similar cut-off radius corresponding to an energy cut-off in
the energy distribution function of the cluster stars, which provided
an even better fit to cluster profiles \citep{King66}.\\

In contrast to that, \citet{Elson87}, and more recently
\citet{Gouliermis10}, found that young massive clusters in the LMC
show an exponential surface density profile without any tidal
truncation at the expected tidal radius. They interpreted these
findings as being due to tidal debris which was expelled at birth from
the clusters and has not had time to disperse yet.\\

Furthermore, \citet{Cote02}, \citet{Carraro07} and \citet{Carraro09}
find the outer halo Milky-Way globular clusters Palomar 13, Whiting 1
and AM 4, respectively, to have a significant excess of stars at the
tidal boundary, which makes any fit to the surface density data very
inaccurate. For all clusters they find the radial surface density
profile, $\Sigma(R)$, to be well represented by a power-law $R^\eta$
with slopes of about $\eta \simeq -1.8$. This excess of stars is
interpreted by the authors as heavy mass loss in a final stage of
dissolution. The same was found for Palomar 5, which is a well studied
MW globular cluster close to the apogalacticon of its orbit
\citep{Odenkirchen03}.\\

Moreover, \citet{McLaughlin05} showed that most globular clusters of
the Milky Way, the LMC and SMC, as well as of the Fornax dwarf
spheroidal, are more extended than can be explained by \citet{King66}
models, and therefore are better represented by \citet{Wilson75}
models. In this context they emphasised the lack of physical
understanding of this phenomenon.\\

On the contrary, \citet{Barmby09} found that of 23 young massive
clusters in M31 most were better fitted by \citet{King66} models as
these clusters do not show extended haloes.\\

As a consequence of this lack of understanding, \citet{Barmby09} asked
in their investigation how robust the physical parameters are which
were derived in such analyses. This they tried to estimate by
analyzing artificial clusters in the same way as the real
observations. A similar analysis has also been performed by
\citet{Bonatto08}, although they tested the analytic formula of
\citet{King62} with artificial observations under various limiting
conditions. Both investigations came to the conclusion that physical
parameters can in principle be well recovered from such idealized mock
observations.\\

But probably the idealized nature of these investigations is
misleading, since both tests were performed with dynamically unevolved
clusters. However, a self-consistent test of deriving physical
parameters from a set of numerical computations of star clusters with
a range of initial parameters has not been performed yet. Just a few
investigations have touched this topic so far by means of numerical
computations (e.g. \citealt{Capuzzo05}, \citealt{Drukier07},
\citealt{Trenti10}). This is due to the fact that fast codes for
globular cluster integration like Fokker-Planck or Monte-Carlo codes
are not able to address this problem properly, as they cannot follow
the evolution of the tidal debris and are restricted to cut-off
criteria in energy or angular momentum space. Moreover, limits in
computational power prevented us from carrying out such investigations
by means of collisional $N$-body codes. But the recent improvements in
computational speed of $N$-body codes and the availability of
accelerator hardware (GRAPEs, GPUs) now allows us to study the
dynamical evolution of star clusters with masses of up to several
times $10^4\msun$.\\

Thus, by computing various star clusters in a range of tidal
conditions over several Gyr we investigate the structure of star
clusters, and in particular the evolution of the region around the
tidal radius - the transition region between the star cluster and the
tidal tails - in terms of velocity dispersion and surface density
profiles. The paper is organized as follows: first, a brief
introduction to the topic of potential escapers is given in
Sec.~\ref{Sec:Pesc} as these stars play a key role in this
investigation. Then a few methodological remarks will be made in
Sec.~\ref{Sec:Method} before we  come to the velocity
dispersion profiles in Sec.~\ref{Sec:VDP} and then to the surface
density profiles in Sec.~\ref{Sec:SDP}. In the last section we will
give a summary with a brief discussion.

\section{Potential escapers}\label{Sec:Pesc}
A major problem in producing velocity dispersion and surface density
profiles from observations is how to disentangle cluster members
from background/foreground stars. By only looking at stars which are
inside a projected, estimated Jacobi radius with similar radial
velocities and colours it is at least possible to distinguish between
the cluster population and field stars. Even here, however, the
question arises whether the choice of the cut in radial velocity
significantly influences the results, especially of the velocity
dispersion profile (see \citealt{Kuepper10b}). Moreover, one has to be
careful with how the Jacobi radius was estimated, i.e. was it evaluated
using a mass estimate and was the mass estimated through the velocity
dispersion? Or was the Jacobi radius estimated by a cut-off radius in
the profile and is the assumption that this cut-off radius is equal to
the Jacobi radius reasonable? What if the cluster is on an eccentric
orbit about the galaxy; how does this influence the cut-off radius?\\

Besides, the observer does not know if the stars in the sample which
was extracted in this way are actually members of the cluster or if
they are already evaporated from the cluster and are now part of its
tidal debris, and just lie in projection within the cluster.\\

But even if the observer could distinguish between stars inside the
Jacobi radius and those that are beyond this radius, it would still be
unclear if the stars in the sample are bound to the cluster or have
already gained enough energy to leave the cluster but just haven't
done so yet. These, so-called, potential escapers will have a
significant influence on both the surface density and the velocity
dispersion profile, if there is a non-negligible fraction of them in
the cluster, as these stars will make the profiles deviate from any
theory which does not take them into account - which most theoretical
approaches do not.\\

In numerical modelling of star clusters we have detailed phase space
information on every single star in the computation. Mainly by
numerical investigations it has been found that the escape process
through which stars escape from a cluster in a constant tidal field,
e.g. on a circular orbit about a galaxy, is divided into two
steps. First the stars get unbound via two-body relaxation, thus on a
relaxation time scale (ignoring constants),
\begin{equation}\label{eq:trel}
t_{rel} \propto N \,t_{cr},
\end{equation}
where $N$ is the number of stars in the cluster and $t_{cr}$ is the
crossing time of the cluster (note that we neglected the slowly
varying Coulomb logarithm in eq.~\ref{eq:trel}, for further details
see e.g. \citealt{Heggie03}). In the second step, these energetically
unbound stars, or potential escapers, with specific energy $E$ higher than some critical escape energy $E_{crit}$, escape from the cluster on an escape time scale which is given by (\citealt{Fukushige00}, equation 9 therein)
\begin{equation}
t_{esc} \propto t_{cr}\,
\left(\frac{E-E_{crit}}{E_{crit}}\right)^{-2},
\end{equation}
where $E_{crit}$ is the critical energy at the Lagrange points, i.e. at the Jacobi radius. From this it
follows that the excess energy can be written as
\begin{equation}
\frac{E-E_{crit}}{E_{crit}} \propto (t_{cr}/t_{esc})^{1/2}.
\end{equation}
On the other hand, the relation
\begin{equation}
\frac{E-E_{crit}}{E_{crit}} \propto (t_{esc}/t_{rel})^{1/2}
\end{equation}
holds, whence we find that
\begin{equation}
t_{esc} \propto (t_{cr}\, t_{rel})^{1/2}.
\end{equation}
Hence, a fraction of stars with excess energy $(E-E_{crit})/E_{crit}$ escapes on a time
scale $t_{esc}$ and the time scale of mass loss, e.g. the time on
which a cluster dissolves, $t_{diss}$, is given by
\begin{eqnarray}
t_{diss} &\propto& t_{esc}
 \,\left(\frac{E-E_{crit}}{E_{crit}}\right)^{-1}\\ & \propto&
 t_{cr}^{1/4}\,t_{rel}^{3/4}\\ & \propto& t_{rel}\,N^{-1/4}
\end{eqnarray}
which was also found by \citet{Baumgardt01}. This result is in
contrast to the `classic' picture where the dissolution times of
clusters scale with the relaxation time (see e.g. \citealt{Binney87}),
i.e.
\begin{equation}
t_{diss} \propto t_{rel},
\end{equation}
and demonstrates the importance of potential escapers on the
dissolution process of star clusters. The remaining question is
whether the population of potential escapers is large enough that it
also has an influence on the velocity dispersion and surface density
profiles. For clusters of a few $10^4$ stars in a constant tidal field
\citet{Baumgardt01} finds about 10-20\% of all stars within the Jacobi
radius to be potential escapers, \citet{Just09} even find the fraction
of potential escapers for similar clusters to be 1/3 of the cluster
population. Moreover, \citet{Baumgardt01} found that for clusters in a
constant tidal field the fraction of potential escapers varies with
the number of stars within a cluster approximately as
$N^{-1/4}$. Thus, this fraction should still be significant for
high-$N$ globular clusters.\\

Star clusters in time-dependent tidal fields have not been
investigated in this respect yet, but as tidal perturbations tend to
increase the energy of the stars in a cluster
(e.g. \citealt{Gnedin99}), the population of potential escapers should
be even larger in such clusters. Thus, potential escapers should have
a significant effect on both the velocity dispersion and the surface
density profiles of all kinds of star clusters.

\section{Method}\label{Sec:Method}
When analyzing velocity dispersion and surface density data of star
clusters,  confusion over nomenclature often arises. Therefore
we strictly stick to the following three terms: \textit{profiles},
\textit{models} and \textit{templates}. The first will be used for the
data, which in our case comes from $N$-body computations but might also
originate from observations. The term \textit{models} will be used for
physically motivated distribution functions such as those proposed by \citet{King66} or
\citet{Wilson75}. The word \textit{templates} covers the analytic
expressions which have been found empirically and which do not
originate from a physical derivation by being solutions of the Collisionless Boltzmann Equation (e.g. \citealt{King62, Elson87,
Lauer95}).  The word template hereby emphasises their artificial
nature.\\

Further confusion arises about the term \textit{tidal radius}. Again
there is a physically motivated and an empirical version. The former
is the radius at which the internal gravitational acceleration equals
the tidal acceleration from the host galaxy (see
e.g. \citealt{Binney87}). This will be named \textit{Jacobi radius} in
our text. The empirical tidal radius is defined through the cut-off
radius of the models and templates which are fitted to the
profiles. We will refer to this cut-off radius as \textit{edge radius}
here and would like to emphasise that Jacobi radius and edge radius
are two different concepts and therefore are not necessarily equal. In
fact, we will show that the two are often significantly different and
that conclusions from a fitted edge radius should be drawn with
caution.\\

Note also that the results of this investigation cannot be easily
scaled to more massive globular clusters, as the relative importance of
tidal features depends on the mass of the cluster, since the mass in
the tidal debris is proportional to the mass-loss rate of the cluster
and the mass-loss rate does not scale linearly with cluster mass, $M$,
but rather with $M^{1/4}$ \citep{Baumgardt03}.\\

Moreover, even though we have some concentrated clusters in our sample
this investigation focuses on less concentrated clusters, with a ratio
of projected half mass radius to Jacobi radius between $0.1 <
R_{hp}/R_J < 0.3$ because \citet{Baumgardt10} showed that clusters
with ratios of $R_{hp}/R_J < 0.05$ cannot  be properly fitted by either
\citet{King62} templates or \citet{King66} models as their ratio
of $R_{hp}/R_J$ is too high. Therefore, our conclusions mainly hold
for less concentrated clusters which \citet{Baumgardt10} named the
\textit{extended cluster population}, and which are likely to be on
the main sequence of cluster evolution \citep{Kuepper08b}, the final
and universal stage of cluster evolution after core collapse, where
the ratio $R_{hp}/R_J$ only depends on the mass of the cluster.\\

In the following sections we will first give some details about our data
set before we explain how we extract the velocity dispersion and
surface density profiles. Thereafter we briefly describe the
analytical templates which are used in this investigation and specify
our fitting method.

\subsection{Data set}\label{Sec:data}
\begin{table}
\begin{minipage}{84mm}
\centering
 \caption{Overview of all computed models. $R_G^{apo}$ gives the
 apogalactic radius of the cluster orbit and $\epsilon$ the
 corresponding eccentricity. The latter is defined as $\epsilon =
 (R_G^{apo}-R_G^{peri})/(R_G^{apo}+R_G^{peri})$ with $R_G^{apo}$ being
 the apogalactic radius of the cluster orbit, and $R_G^{peri}$ being
 the perigalactic radius.  $R_{hp}/R_J$ gives the
 initial concentration of the model, whereas $R_h$ is the initial
 half-mass radius. The clusters are divided into three categories:
 extended clusters in a constant tidal field (A0-D0), extended
 clusters in time-dependent tidal fields (A1-D3) and compact clusters
 (A0c-D0c). Note that each model has initially $N=64k$ stars, a lower
 stellar mass limit of $0.1\msun$ and an upper limit of $1.2 \msun$. }
\label{table1}
\begin{tabular}{lcccc}
\hline Name & $R_G^{apo} \, [\mbox{kpc}]$ & $\epsilon$ &
 $\frac{R_{hp}}{R_J}$ & $R_h$ [pc]\\ \hline A0& 4.25 & 0.00 & 0.15 &
 5.8 \\ B0& 8.50 & 0.00 & 0.15 & 8.8\\ C0& 12.75 & 0.00 & 0.15 &
 11.7\\ D0& 17.0 & 0.00 & 0.15 & 14.3\\ \hline A1/A2/A3 & 4.25 &
 0.25/0.50/0.75 & 0.15 & 5.8\\ B1/B2/B3 & 8.50 & 0.25/0.50/0.75 & 0.15
 & 8.8\\ C1/C2/C3 & 12.75 & 0.25/0.50/0.75 & 0.15 & 11.7\\ D1/D2/D3 &
 17.0 & 0.25/0.50/0.75 & 0.15 & 14.3\\ \hline A0c & 4.25 & 0.00 & 0.08
 & 3.0\\ B0c & 8.50 & 0.00 & 0.06 & 3.0\\ C0c & 12.75 & 0.00 & 0.05 &
 3.0\\ D0c & 17.0 & 0.00 & 0.04 & 3.0 \\
\end{tabular}
\end{minipage}
\end{table}
The velocity dispersion and surface density profiles are mainly taken
from our recent investigation of star clusters and their tidal tails
\citep{Kuepper09}. For this analysis we computed a comprehensive set
of star clusters on various orbits about a Milky-Way potential.\\

In addition to the 16 clusters taken from \citet{Kuepper09}, we computed 4
concentrated clusters which are of specific interest for this
investigation. All clusters were set-up using the publicly available code \textsc{McLuster}\footnote{\tt www.astro.uni-bonn.de/\~{}akuepper/mcluster/mcluster.html}  (K\"upper et al., in prep.), and integrated over time with the $N$-body code \textsc{NBODY4} \citep{Aarseth03} on the
\textsc{GRAPE-6A} computers at AIfA Bonn \citep{Fukushige05}. The code
was modified so that the Milky-Way potential suggested by
\citet{Allen91} could be used. Due to its analytic form this potential
is useful for calculating the Jacobi radius, $R_J$, of the star
clusters. Throughout the computations $R_J$ was evaluated using
equation 12 of \citet{Kuepper09}, i.e.
\begin{equation}\label{eq:R_J}
R_J = \frac{GM}{\Omega^2- \partial^2\Phi/\partial R_G^2},
\end{equation}
where $G$ is the gravitational constant, $M$ is the cluster mass,
$\Omega$ is the angular velocity of the cluster on its orbit about the
galaxy, and $\partial ^2\Phi/\partial R_G^2$ is the second derivative
of the gravitational potential of the galaxy with respect to the
galactocentric radius, $R_G$. We always determine the Jacobi radius
iteratively, assuming first a mass of the cluster, then applying
eq. \ref{eq:R_J}. With this estimate we determine the mass of the
cluster again, counting all masses around the cluster centre within
$R_J$, and compute the Jacobi radius once more, and so on until the
value of $R_J$ converges.\\

The clusters were set up using a Plummer density distribution with a
ratio of half-mass radius to Jacobi radius of $R_h/R_J = 0.2$, i.e. a
ratio of projected half-mass radius to Jacobi radius of $R_{hp}/R_J =
0.15$. This choice resembles the extended globular clusters of the
Milky Way, which are less concentrated, and stand in contrast to the compact
clusters, which are deeply embedded within their Jacobi radius and for
which tidal influences are rather negligible \citep{Baumgardt10}.\\

All clusters had initially 65536 stars which were drawn from the
canonical IMF\footnote{The canonical IMF has a slope of $\alpha_1 =
1.3$ for stellar masses $m = 0.08 - 0.5 \msun$, and the Salpeter slope
$\alpha_2 = 2.3$ for $m > 0.5 \msun$.} \citep{Kroupa01} ranging from
$0.1\msun$ to $1.2 \msun$, resulting in a total initial mass of about
$20\,000 \msun$. All clusters were modelled without stellar evolution
and without primordial binaries. They were computed for a time span of
4 Gyr, if they did not dissolve before reaching this age. An overview
of all clusters in our sample is given in Tab.~\ref{table1}.\\

For all clusters we produced two-dimensional snapshots of the stars
projected onto the orbital plane every 10 Myr; thus, as each of the 20
clusters is modelled over 4 Gyr, we have about $20\times 400 \sim
10^4$ representations of star clusters. In each snapshot we determine
the centre of the cluster stars within the Jacobi radius following
\citet{Casertano85} and bin the stars around this centre in 50 annuli
of equal logarithmic width between 1.0 pc and 500 pc to produce the
velocity dispersion and surface density profiles. By going out to 500
pc we are able to show not only the region close to the Jacobi radius
but also a considerable part of the tidal tails, which have not been
investigated in this respect yet, but show an interesting, variable
behaviour, depending on the cluster orbit, which can even influence the
velocity dispersion and surface density within the Jacobi radius (see
\citealt{Kuepper09}). For most of the investigation, however, we concentrate on
the inner 100 pc.

\subsection{Velocity dispersion profiles}
In each annulus the velocity dispersion along the line-of-sight (los),
$\sigma$, is determined using
\begin{equation}
\sigma^2 = \frac{1}{N} \sum_{i=1}^N\left(v_i - \overline{v}\right)^2 =
\overline{v^2}-\overline{v}^2,
\end{equation}
where $v_i$ is the velocity of the $i$-th star along the
line-of-sight, $\overline{v}$ is the mean los velocity of the $N$
stars in the annulus, and $\overline{v^2}$ is their mean squared
los velocity. In addition, we estimate an uncertainty for each velocity
dispersion,
\begin{equation} 
\Delta \sigma^2 = \frac{1}{N^2 \sigma^2} \sum_{i=1}^N \left( \left(v_i
- \overline{v} \right)^2 - \sigma^2 \right)^2,
\end{equation}
which originates from a Taylor expansion of the standard deviation of
the velocity dispersion.\\

The velocity dispersion is first measured by taking into account all stars in the
snapshot, independent of their projected distance from the cluster
centre, which is what an ideal observer would see. Then
it is measured for only the stars within the Jacobi radius, i.e. $r
\leq R_J$. In this way we will investigate the effect of
foreground/background stars, i.e. stars in the tidal debris, on the
observed velocity dispersion. Finally we measure $\sigma$ taking only
the bound stars into account to test if energetically unbound stars
have a significant influence on the observed velocity dispersion.\\

For this purpose we have to search for potential escapers and remove
them from the sample. Thus, bound stars are defined as stars which are
inside the Jacobi radius and which have a Jacobi energy smaller than
\begin{equation}
E_{crit} = -\frac{3GM}{2R_J},
\end{equation}
which is the critical energy at the Lagrange points, i.e. at the Jacobi
radius. We therefore compute the Jacobi energy, $E$, of each star with
mass $m$ in a co-rotating reference frame, according to the near-field
approximation which  can be found in \citet{Spitzer87}:
\begin{equation}
E = \frac{mv^2}{2} - \frac{GMm}{r} +
\frac{m}{2}\Omega^2\left(z^2+3x^2\right),
\end{equation}
where the first term is the kinetic energy of the star, the second
term is its energy in the gravitational field of the cluster with mass
$M$ at the star's radius $r$, and the third term is a combination of
the centrifugal and tidal potentials. In this last term $z$ gives the
distance of the star from the cluster centre perpendicular to the
cluster's orbital plane in the galaxy, whereas $x$ denotes the
coordinate of the star relative to the cluster centre, along the axis
pointing to the galactic centre (for a detailed description see
e.g. \citealt{Fukushige00}). We are well aware of the fact that this
approximation holds only for nearly circular cluster orbits in a
point-mass galactic field, and take it only as a first-order
approximation here.\\

Note that we remove binaries from our data and replace them by their
centre-of-mass particles, since we want to focus on the effect of
potential escapers on the velocity dispersion. In observational data,
binaries, of course, can play a major role, as has recently been shown
by \citet{Gieles09}, \citet{Kouwenhoven09} and \citet{Kuepper10b} but only for clusters with a velocity dispersion of less than about 5-10 km/s.

\subsection{Surface density profiles}
In each annulus the surface density, $\Sigma(R)$, is determined by
counting the number of stars, $N$, in the given (projected) radial
range $[R-\Delta R, R]$, and dividing by the surface area, $A$, of the
given annulus, i.e.
\begin{equation}
\Sigma(R) = \frac{N}{A} = \frac{N}{2\pi \left(R^2-(R-\Delta
R)^2\right)}.
\end{equation} 
Moreover, for each annulus an uncertainty is estimated by the square
root of the number of stars in the annulus divided by its surface
area,
\begin{equation}
\Delta \Sigma(R) = \frac{\sqrt{N}}{2\pi \left(R^2-(R-\Delta
R)^2\right)}.
\end{equation} 
First the surface density profiles are evaluated by taking into
account all stars in each sample, and then additionally only the bound
stars, to show where and in which way potential escapers influence the
profile.
 
\subsection{Analytical templates}\label{sec:templates}
\begin{table}
\begin{minipage}{84mm}
\centering
 \caption{Overview of the templates used in this investigation. $n_f$
 gives the number of free parameters, where the +1 denotes an
 additionally assumed constant background. As each template was
 originally designed with the focus on specific components of star
 clusters, this table shows for each template  which of the
 three basic components it can handle: core, bulk and tidal debris.}
\label{table2}
\begin{tabular}{lcccc}
\hline Name & $n_f$ & core & bulk & tidal debris\\ \hline King & 3+1 &
   flat & separated & n/a\\ EFF & 3+1 & flat & not sep. & power-law\\
   Nuker & 5+1 & cuspy & not sep. & power-law\\ gen. Nuker & 8 & cuspy
   & separated & power-law\\ KKBH & 6+1 & cuspy & separated &
   power-law\\
\end{tabular}
\end{minipage}
\end{table}
\begin{figure}
\includegraphics[width=84mm]{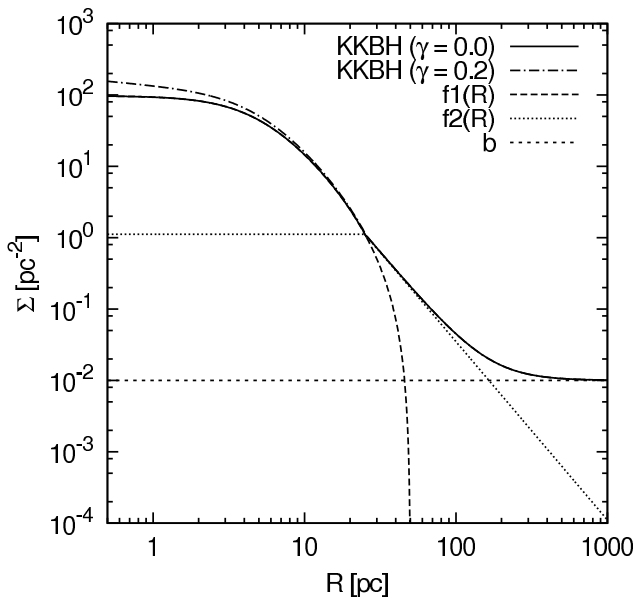}
  \caption{Sketch of the KKBH template for two different values of the
  core slope $\gamma$. Also shown are the two sub-components
  for the bulk ($f1(R)$, eq. \ref{eq:ETC1}), which resembles the King
  (1966) template, and the tidal debris ($f2(R)$, eq.~\ref{eq:ETC2}),
  which resembles the \citet{Elson87} template. The break radius
  between the two components is at $\mu R_t$, where $\mu$ is found to
  be 0.5 in most cases, i.e. for radii larger than 50\% of the edge
  radius the surface density profiles of the clusters follow a
  power-law. More flexibility for large radii is added to most
  templates through a constant background $b$. }
  \label{ETC}
\end{figure}
Even though physically motivated models like \citet{King66} or
\citet{Wilson75} have often proved to more generally represent
observations of globular clusters than analytical templates
(e.g. \citealt{McLaughlin05}), they have one decisive flaw: there is
no simple analytical description which can be quickly fitted to star
cluster profiles. Thus, due to their simplicity, analytical templates
are still in use, especially for larger data sets of extra-Galactic
clusters.\\

Moreover, analytical templates have the advantage that they can be
easily set up and extended, according to the needs of the
corresponding investigation. We found that star clusters  consist
basically of three parts: core, bulk and tidal debris. The templates
which have been set up in the past were adjusted according to the different foci
of the corresponding investigations. For example, \citet{King62} was
mainly interested in the bulk of the cluster, and so the template mainly
consists of a bulk and just a flat core, whereas \citet{Elson87}
focused on the tidal debris of young clusters, and so their template
has a flat core, no separated, explicitly defined bulk but a power-law
distribution of
tidal debris. Therefore, we will  study some existing templates in a
systematic way to show their advantages and their flaws, and finally
design a more general template. An overview of all templates is given
in Tab.~\ref{table2}.\\

The `classic' templates we use, their abbreviations for the rest of
the text, and their fit parameters are the following:
\begin{itemize}
\item \textbf{King} denotes the empirical fitting formula which was
devised by \citet{King62} and which works surprisingly well for many
extended globular clusters (i.e. $R_{hp}/R_J > 0.05$) and open clusters. It has the form
\begin{equation}
f(R) = k \left[ \frac{1}{\sqrt{1+\left(R/R_c\right)^2}} -
\frac{1}{\sqrt{1+\left(R_t/R_c\right)^2}} \right]^2 + b
\end{equation}
for $R<R_t$ and $f(R) = b$ for $R\ge R_t$, where $k$ is a scale
factor, $R_c$ is a core radius, and $R_t$ is often denoted as the
tidal radius - in our nomenclature it is the edge radius, the limiting
radius between the bulk of a cluster and its tidal debris. The King
template therefore consists of a flat core and a bulk, but has no term
for the tidal debris. Note that we allow for a constant background $b$
instead, as would be done in real observations even though we have no
real background but only the tidal debris of the clusters. We found this
constant to significantly improve the quality and stability of the
fit.\\
\item \textbf{EFF} is the empirical template used  in the work of
\citet{Elson87} on young LMC clusters of about the same mass as the
clusters in our sample. They found that the outer parts of those
clusters were poorly fitted by the King template and adopted this
template which has no edge radius but falls off like a power-law:
\begin{equation}
f(R) = k \left[ 1+ \left(\frac{R}{R_c}\right)^2\right]^{-\eta/2} + b,
\end{equation}
where $k$ is a scale factor, $R_c$ is a core radius, $b$ is a constant
background and $\eta$ is the slope of the template for radii much
larger than the core radius. The EFF template was designed to fit
observations of clusters with pronounced tidal debris, and therefore
consists of a flat core and an unseparated bulk and tidal debris which
are represented by a single power-law.\\
\item \textbf{Nuker} denotes the template which was adopted by \citet{Lauer95} for
elliptical galaxies and which has also occasionally been used for star
clusters. It is also a power-law but shows more flexibility than EFF
and can also fit a cluster with a non-flat core, e.g. a core-collapsed
cluster or a cluster with a massive central black hole. It has the
form
\begin{equation}\label{eq:nuker}
f(R) = k\,2^{\frac{\eta-\gamma}{\alpha}}
\left[\frac{R}{R_c}\right]^{-\gamma} \left[1 +
\left(\frac{R}{R_c}\right)^\alpha\right]^{-\frac{\eta-\gamma}{\alpha}}+b,
\end{equation}
where $k$ is a scale factor, $R_c$ is a core radius and $b$ is a
constant background. $\gamma$ gives the power-law slope inside $R_c$,
whereas $\eta$ gives the slope for radii larger than the break radius
$R_c$. With the factor $\alpha$, the smoothness of the transition
between the two slopes is defined. Thus, the Nuker template consists
of a flexible core and a power-law tidal debris.\\
\end{itemize}

Additionally, we set up two templates which have a flexible core plus
an edge radius, i.e. a well-defined bulk, but can also fit the region
beyond the edge radius in a more sophisticated way than by just a constant
background.
\begin{itemize}
\item Generalised Nuker, \textbf{gen.~Nuker}, is the same as Nuker but
allows for a different power-law slope for radii larger than a second
break radius. The template has been introduced by
\citet{vanderMarel09} to get a smooth representation of the surface
density profile of $\omega$ Cen. It has the form
\begin{eqnarray}\label{eq:gennuker}
f(R) &=& k\,2^{\frac{\eta-\gamma}{\alpha}}
\left[\frac{R}{R_c}\right]^{-\gamma} \left[1 +
\left(\frac{R}{R_c}\right)^\alpha\right]^{-\frac{\eta-\gamma}{\alpha}}\times\nonumber\\
& &\left[1 +
\left(\frac{R}{R_t}\right)^\delta\right]^{-\frac{\epsilon-\eta}{\delta}},
\end{eqnarray}
where the parameters are the same as in Nuker, though without the
constant background, but instead with an additional second break
radius, $R_t$, a slope for radii larger than this radius, $\epsilon$,
and a factor $\delta$ which determines the smoothness of the
transition between $\epsilon$ and $\eta$. In their investigation of
$\omega$ Cen, \citet{vanderMarel09} found that the two break radii in
this template roughly correspond to the core and Jacobi radius of the
cluster. In our nomenclature their template consists of a flexible
core, a bulk and a tidal debris, which are separated by two break
radii: the core radius and the edge radius.\\

\item Finally, we set up our own template which is tailored to the
purpose of this investigation. \textbf{KKBH} is based on the King
template but modified in two steps: first we modified the King
template in much the same way as the Nuker template is modified with
respect to EFF, i.e. we allow it to have a power-law cusp in the core,
but without changing the behaviour in the outer parts of the template:
\begin{eqnarray}
f(R) &=& k \left[\frac{R/R_c}{1+R/R_c}\right]^{-\gamma}
\times\nonumber\\ & & \left[ \frac{1}{\sqrt{1+\left(R/R_c\right)^2}} -
\frac{1}{\sqrt{1+\left(R_t/R_c\right)^2}} \right]^2 + b
\end{eqnarray}
for $R<R_t$ and $f(R) = b$ for $R\ge R_t$, with the same parameters as
King plus an additional core power-law slope $\gamma$. For $\gamma =
0$ it gives the original King template. This template therefore
consists of a flexible core and a bulk but still has no tidal debris
term, yet. \\

For this reason we modified it further with an additional extra-tidal component in
 the form of a power-law slope. For the core and the inner part of the bulk
of the cluster it behaves like the previous template, but at a radius
which is a fraction $\mu$ of the edge radius the template changes
abruptly into a power-law, and thus behaves like EFF (see
Fig.~\ref{ETC}). It is given by
\begin{eqnarray}\label{eq:ETC1}
f_1(R) &=& k \left[\frac{R/R_c}{1+R/R_c}\right]^{-\gamma}
\times\nonumber\\ & & \left[ \frac{1}{\sqrt{1+\left(R/R_c\right)^2}} -
\frac{1}{\sqrt{1+\left(R_t/R_c\right)^2}} \right]^2
\end{eqnarray}
for radii smaller than $\mu\,R_t$, and
\begin{equation}\label{eq:ETC2}
f_2(R) = f_1(\mu\,R_t) \left[1 +
\left(\frac{R}{\mu\,R_t}\right)^{64}\right]^{-\eta/64}
\end{equation}
for $R\ge\mu\,R_t$, where the parameters are the same as for the
previous template, except for the new ones in the function $f_2(R)$. The exponent
64 in $f_2(R)$ causes the template to change abruptly into the
power-law slope of $\eta$ at a fraction $\mu$ of the edge radius. With
an additional constant background, $b$, the complete function looks as
follows when using ternary operators
\begin{equation}
f(R) = \left( R < \mu R_t \quad ? \quad f_1(R) + b \quad : \quad
f_2(R) + b\right)
\end{equation}
(meaning: if $R$ is smaller than $\mu R_t$ use $f_1(R)+b$, else use
$f_2(R)+b$). Defined in this way, the KKBH template has a flexible
core, a well-defined bulk and a power-law tidal debris (see
Fig.~\ref{ETC}). A brief explanation on how to use the KKBH template with
gnuplot is given in Appendix A.  \\

The three additional parameters of KKBH may not be useful in all
applications but are motivated through the following facts:
\begin{itemize}
\item Mass segregation or the presence of an intermediate-mass black
hole can have a significant influence on the slope of the core
profile. This cannot be fitted and thus quantified by the original
King template, and therefore we add the parameter $\gamma$ to the
template.
\item In most applications the star counts in the outer parts of
surface density profiles drop quickly, while their uncertainties grow,
so that a template, and hence the edge radius, is mostly fitted by
the inner profile \citep{Baumgardt10}. Hence, by allowing a smooth
template for the inner profile and an independent power-law for the
outer profile we decouple the two parts and look for the transition
point between the two, which is set by the second additional parameter
$\mu$.
\item Potential escapers and background/foreground stars as well as
the tidal debris of the cluster influence the template fits in the
outer parts of profiles \citep{Cote02, Carraro07, Carraro09}. Thus,
the tidal debris is not reflected properly by templates like King, and so the
results of
 template fitting are not independent of the presence of debris. Furthermore, by
measuring the slope of the surface brightness profile outside the edge
radius we aim to be able to  deduce information on the orbital phase of the
cluster, and therefore we add the parameter $\eta$.
\end{itemize}
\end{itemize}

\subsection{Fitting method}
We fit the above templates to the data by finding a minimum in
$\chi^2$ using a nonlinear least-squares Marquardt-Levenberg
algorithm.  Here we define $\chi^2$ as a measure of the differences between
the template, $f(R)$, and the data, $\Sigma(R)$, weighted by the
uncertainties in the surface density data, $\Delta \Sigma(R)$, summed
up over all data points, i.e.
\begin{equation}
\chi^2 = \sum_i {\frac{\left( \Sigma(R_i)-f(R_i) \right)^2}{\Delta
\Sigma(R_i)^2}}.
\end{equation}
To compare the fits to each other and to account for the different
complexity of the templates we calculate a reduced $\chi^2$ which is
given by the above $\chi^2$ divided by the number of degrees of
freedom, $n$,
\begin{equation}
\chi_{red}^2 = \frac{\chi^2}{n},
\end{equation}
where $n$ is given by the number of data points, $n_d$, minus the
number of parameters of the given template, $n_f$, minus 1 (as one
parameter can be fixed by the mean value of the data points), i.e.
\begin{equation}
n=n_d-n_f-1.
\end{equation}
The resulting values of $\chi^2_{red}$ should ideally be close to
one, as too low  a value would imply that the template is too complex
for the underlying data.\\

Each template is fitted to each cluster for five different ranges of
radii to see if the goodness of the fit changes for different
coverages of the cluster and its debris, which also implies a
differing number of data points. Therefore we choose fixed radial
intervals of 1-25 pc, 1-50 pc, 1-100 pc, 1-200 pc and 1-500 pc. Since
the number of data points between 1-500 pc is fixed to 50 for all
clusters, we get [25, 31, 37, 42, 50] data points in the given radial
ranges. As the clusters have the same initial mass but orbit the
galaxy at different galactocentric radii, these five intervals will
cover quite different parts of the clusters. For instance a cluster of
$20\,000 \msun$ has a Jacobi radius of about [27, 40, 52, 63] pc for a
circular orbit at [4.25, 8.5, 12.75, 17] kpc (eq.~\ref{eq:R_J}), respectively.

\section{Results}
\subsection{Velocity dispersion profiles}\label{Sec:VDP}
\subsubsection{Constant tidal fields}
\begin{figure*}
\includegraphics[width=55mm]{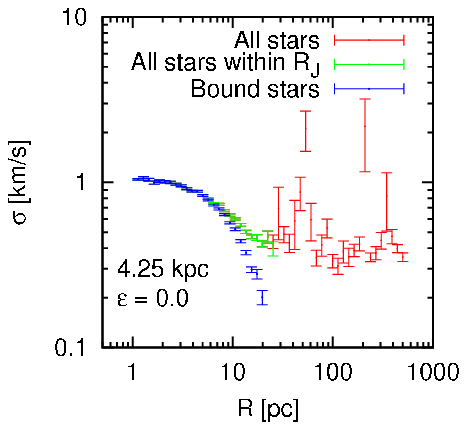}
\includegraphics[width=55mm]{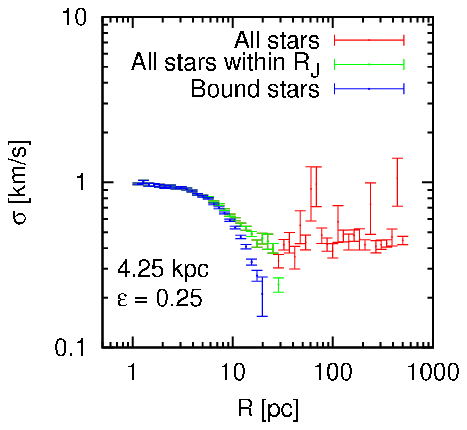}
\includegraphics[width=55mm]{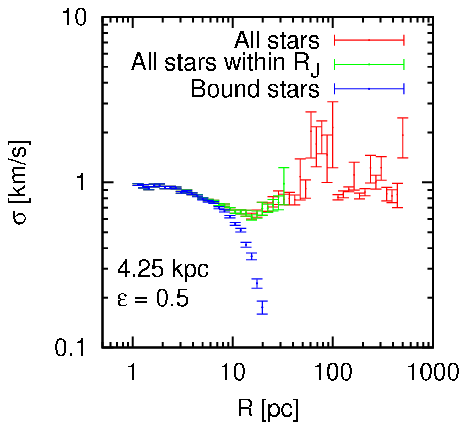}\\
\includegraphics[width=55mm]{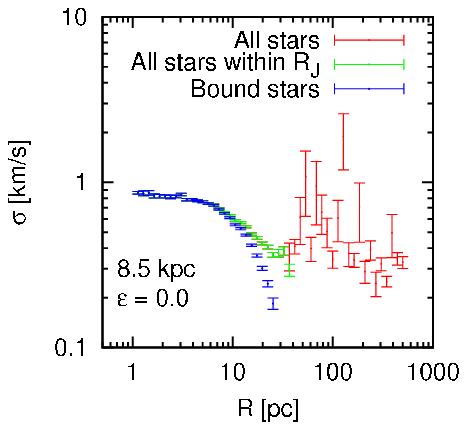}
\includegraphics[width=55mm]{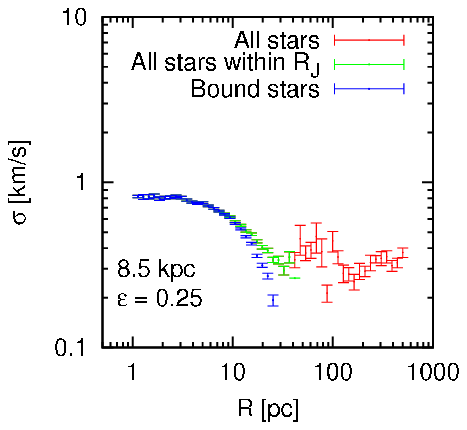}
\includegraphics[width=55mm]{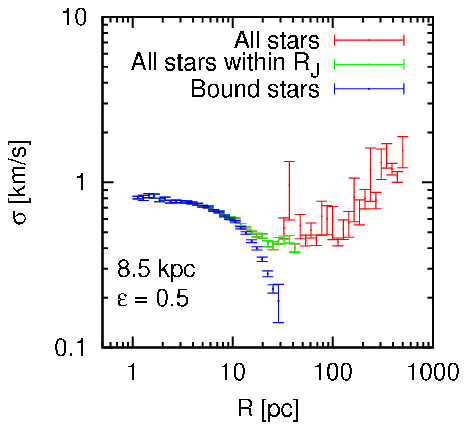}\\
\includegraphics[width=55mm]{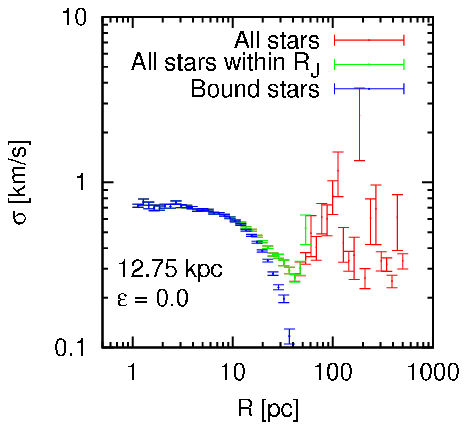}
\includegraphics[width=55mm]{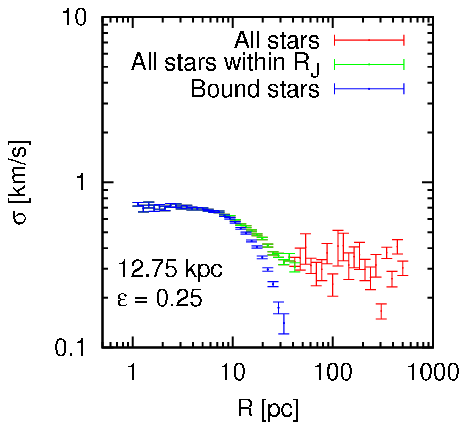}
\includegraphics[width=55mm]{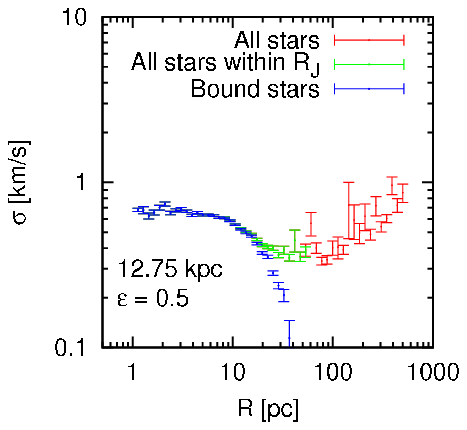}\\
\includegraphics[width=55mm]{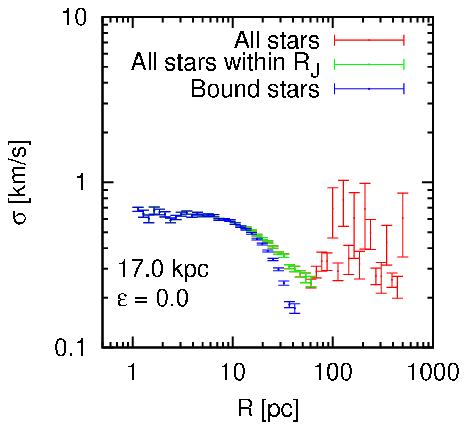}
\includegraphics[width=55mm]{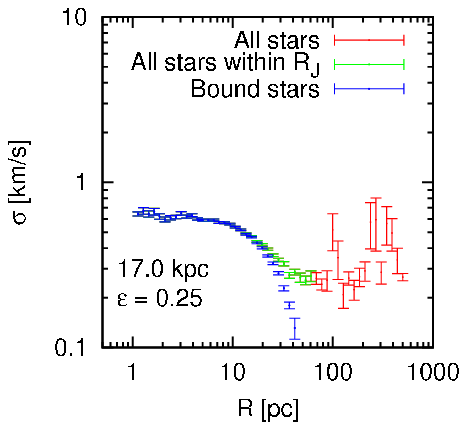}
\includegraphics[width=55mm]{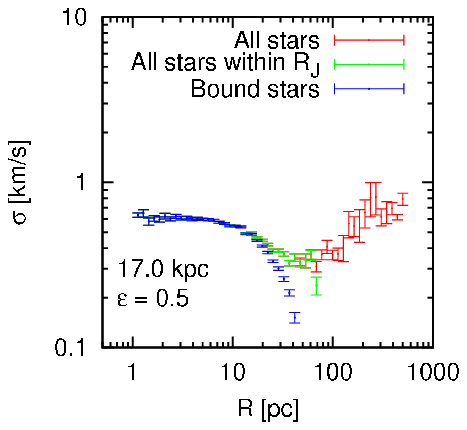}\\
  \caption{Velocity dispersion profiles of 12 different clusters
  (A0-2, B0-2, C0-2, D0-2). The apogalactic distance and eccentricity of
  the clusters' orbits are given in the graphs. At the given snapshot
  each cluster is at apogalacticon and has a mass of about
  $15\,000\msun$. The corresponding Jacobi radius can be deduced by the
  outermost green data point. Two effects are visible: first, for
  increasing galactocentric distance the extension of the cluster, and
  hence also the flat part inside the core radius, both grow as the Jacobi
  radius grows; this results from setting up the cluster half-mass
  radii to scale with $R_J$ (Sec. \ref{Sec:data}). This also implies a
  lower central velocity dispersion. Second and more important, for
  increasing eccentricity the effect of potential escapers (green
  points vs. blue points) on the flatness of the velocity dispersion
  profile outside the core radius grows. Corresponding surface density
  profiles are given in Figs.~\ref{sdp8500} \& \ref{sdp8500ecc}.}
  \label{vdps}
\end{figure*}
A typical sample of velocity dispersion profiles is shown in
Fig.~\ref{vdps}. In each graph the velocity dispersion is plotted i)
for all stars in the sample (red), ii) for all stars that lie within
the Jacobi radius (green), and iii) for all bound stars (blue).\\

The first column shows the clusters in a constant tidal field, i.e. on
circular orbits. Comparing the graphs for the different galactocentric
distances shows that, as expected, the clusters at larger
galactocentric distances are more extended. Since they all have the
same mass of about $15\,000 \msun$ in the snapshots, the central
velocity dispersion decreases with $R_G$, whereas the size of the flat
part within the core radius increases.\\

Each of the clusters in a constant tidal field has a steadily declining
velocity dispersion within the Jacobi radius (green/red), although the
clusters in the stronger tidal fields already show an onset of a
flattening of the profile at the Jacobi radius. Beyond the Jacobi
radius the velocity dispersion fluctuates strongly as the number statistics in
these outer bins is quite low and a single fast escaper may dominate
the velocity dispersion in the corresponding bin (red).\\

From the differently coloured curves we can see that the stars which
lie within the projected Jacobi radius, but actually are in the
foreground or background of the cluster within its tidal debris,
hardly influence the velocity dispersion of the clusters (green). Only
in some of the graphs can one make out a small contribution in the very
outermost bins; everywhere else the green data points lie on top of
the red ones.\\

By contrast, potential escapers have a significant influence
on the profiles (green data vs. blue data). By definition the velocity
dispersion of the bound stars approaches zero at the Jacobi radius,
where the tidal forces are equal to the gravitational acceleration of
the cluster. But we see a significant influence already at about 50\%
of the Jacobi radius.

\subsubsection{Time-dependent tidal fields}
\begin{figure*}
\includegraphics[width=55mm]{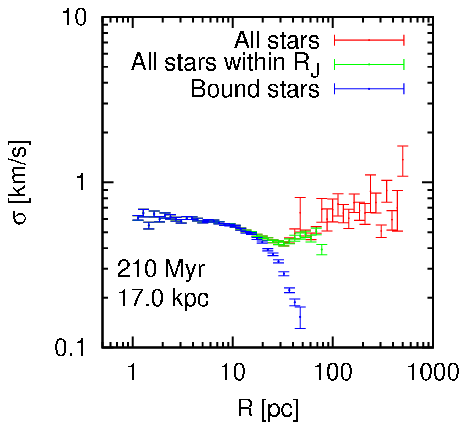}
\includegraphics[width=55mm]{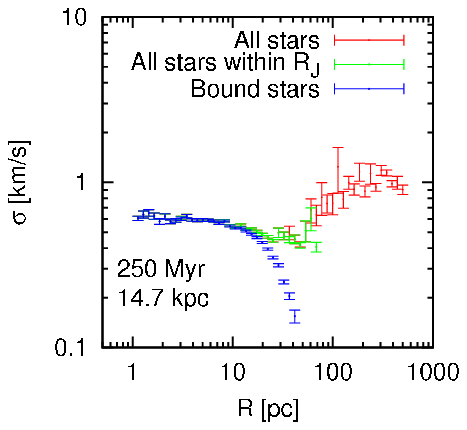}
\includegraphics[width=55mm]{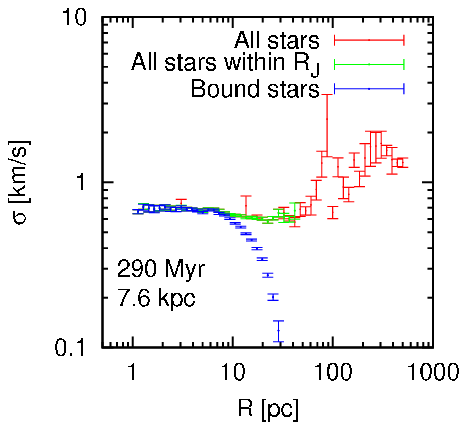}\\
\includegraphics[width=55mm]{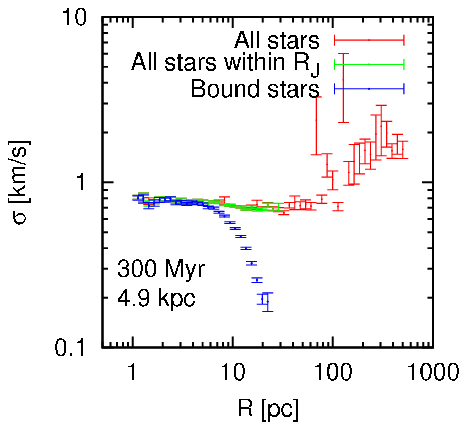}
\includegraphics[width=55mm]{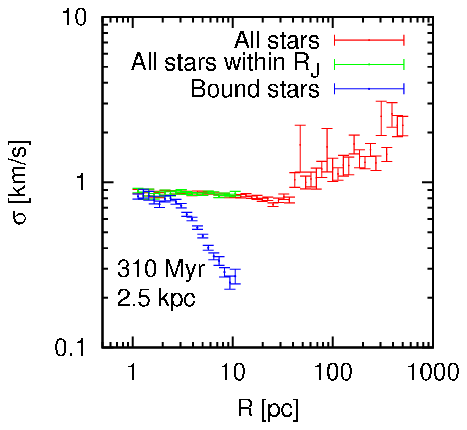}
\includegraphics[width=55mm]{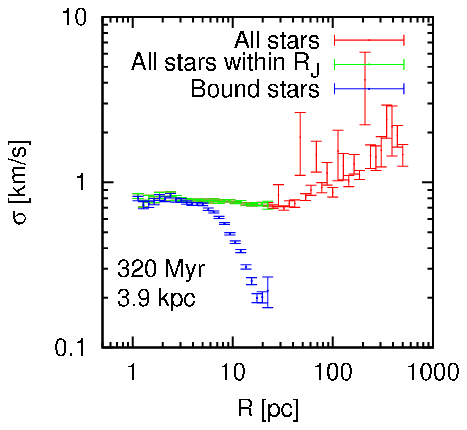}\\
\includegraphics[width=55mm]{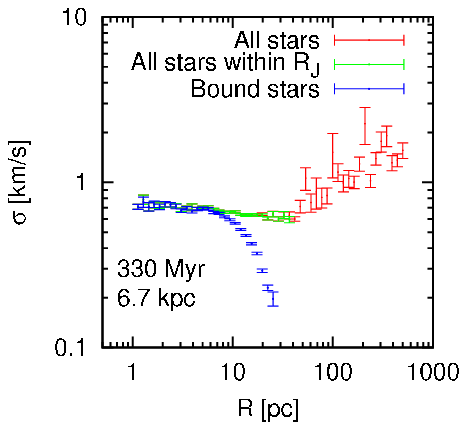}
\includegraphics[width=55mm]{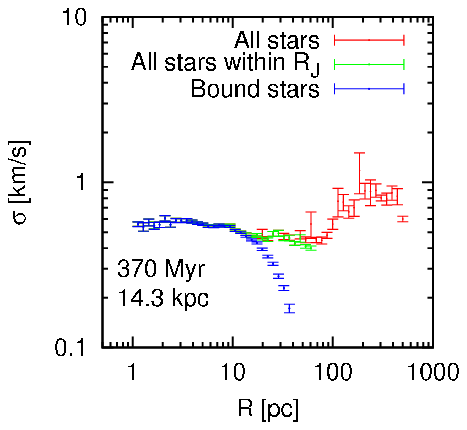}
\includegraphics[width=55mm]{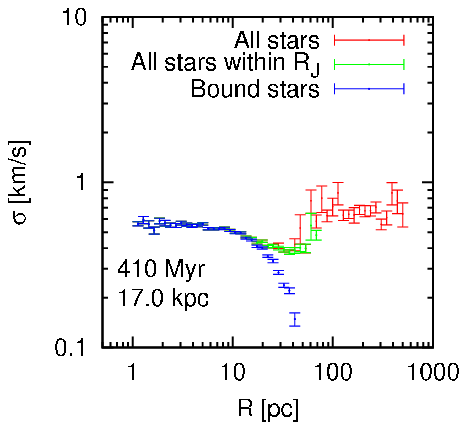}\\
  \caption{Velocity dispersion profiles for the cluster D3, which is on
  an eccentric orbit with eccentricity 0.75 and an apogalactic
  distance of 17 kpc. The snapshots show the evolution of the profile
  with time during one revolution about the galactic centre, from
  apogalacticon to apogalacticon. In each graph the corresponding
  simulation time and galactocentric radius are given. In the middle
  graph the cluster is at perigalacticon. Here the Jacobi radius is at
  its minimum ($\sim$10 pc) and most of the cluster is unbound. For a
  large fraction of the orbit the velocity dispersion profile is
  approximately flat. Corresponding surface density profiles are given
  in Fig.~\ref{sdp17000}.}
  \label{vdpse075}
\end{figure*}
If we take a look at columns 2 and 3 of Fig.~\ref{vdps} we can see the
effect of time-dependent tidal fields on the velocity dispersion
profiles. With increasing eccentricity the extent and onset of the
population of potential escapers stays the same ($\sim 0.5 R_J$),
since in the snapshots all clusters are at apogalacticon. But from the
green curve we can see that the energy of the potential escapers
increases, which leads to a flattening of the velocity dispersion
profiles. This is in agreement with findings by \citet{Gnedin99} who
showed that stars at large radii are most affected by tidal
perturbations like disk shocks or pericentre passages and that stars,
on average, gain energy through such events. Furthermore, it supports
the assumption of \citet{Kuepper09} that stars do not escape from the
cluster immediately after a tidal perturbation, but can remain in the
cluster for several cluster orbits about the galactic centre.\\

In Fig.~\ref{vdpse075} we show a time series of a cluster on an orbit
with an eccentricity of 0.75 and an apogalactic radius of 17 kpc. The
time series starts 210 Myr after the beginning of the computations,
where the cluster is just about to finish its first revolution about
the galactic centre and has a mass of $16\,200\msun$. At 310 Myr
(middle panel) the cluster is at perigalacticon with a mass inside the
Jacobi radius of $7\,700\msun$, and at 410 Myr it is again at
apogalacticon with $13\,500\msun$ left.\\

We see that for the largest part of the orbit, from 250 Myr to 370 Myr
($R_G < 15$ kpc) the velocity dispersion profile appears to be more or
less flat. For the perigalactic part of the orbit (300 Myr - 320 Myr,
$R_G < 5$ kpc) a good fraction of the cluster stars even lies beyond
the Jacobi radius (about 50\% of the apogalactic cluster
mass). Nevertheless, even after such a dramatic pericentre passage the
cluster's velocity dispersion profile returns to nearly its original
shape.

\subsubsection{Concentrated clusters}
\begin{figure}
\includegraphics[width=84mm]{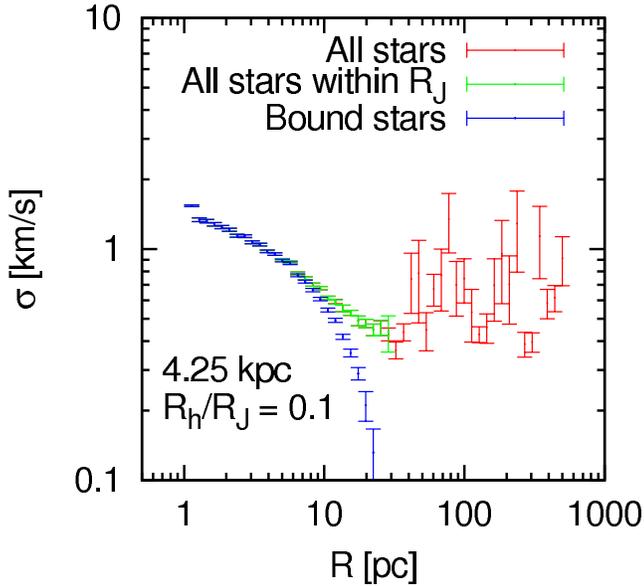}
  \caption{Velocity dispersion profile for the concentrated cluster
  with a galactocentric radius of 4.25 kpc (A0c). At the time of the
  snapshot the cluster has a mass of about $15\,000\msun$ and a ratio
  of projected half-mass radius to Jacobi radius of 0.08. Compared to
  the other clusters at the same stage of evolution, the velocity
  dispersion rises steeply in the centre due to core collapse, but
  shows a similar behaviour in the outer parts. A corresponding
  surface density profile is given in Fig.~\ref{sdpcc}.}
  \label{vdpcc}
\end{figure}
In Fig.~\ref{vdpcc} the velocity dispersion profile for a cluster at
4.25 kpc is shown.  It had an initial (3-dimensional) half-mass
radius of 3 pc, and is thus what \citet{Baumgardt10} would classify as a
compact cluster. At the time of the snapshot the cluster has
$15\,000\msun$ and still has a ratio of projected half-mass radius to
Jacobi radius $R_{hp}/R_J = 0.08$. The profile does not show a
well-defined core in the range of the data points, as all other
profiles do, but rises steeply in the centre. This is a clear
indication of the fact that the cluster is near core collapse. Even
so, there is no significant difference to the other clusters with
respect to the potential escapers.

\subsubsection{Conclusions}
From the velocity dispersion profiles studied in this section we can
conclude that the population of potential escapers has a significant
influence on the kinematical structure of a star cluster. The
influence of this population on the velocity dispersion profiles grows
with increasing eccentricity of the cluster orbit. In such cases an
observer would have to be very cautious when deriving a dynamical mass
out of velocity dispersion measurements as it would tend to be
overestimated,  the possible degree of overestimation depending on
the cluster characteristics such as its orbit and its velocity
dispersion.\\

Furthermore, we saw that the velocity dispersion profile deviates from
the expected Keplerian profile at a radius of about half the Jacobi
radius. Since most globular clusters in the Milky Way experience a
gravitational acceleration which is about the size of $
GM_{Gal}/R_{Gal}^2 \simeq 1.2 \times 10^{-8}\mbox{cm s}^{-2} \simeq
a_0 $ one may easily be misled to deduced deviations from Newtonian
gravity (e.g. \citealt{Scarpa07}). To reduce the influence of
potential escapers, velocity dispersion measurements should be made
within 50\% of the Jacobi radius. Nevertheless, extra caution has to
be taken for clusters on eccentric orbits.\\

Moreover, we can see the general structure of star clusters already in
the velocity dispersion profiles. There is a core which can be flat or
cuspy. Then there is a bulk of bound stars whose limiting radius is
somewhat smaller than the Jacobi radius, and finally there is the
tidal debris whose onset lies within the Jacobi radius as a
consequence of potential escapers. Most of this cluster structure is
temporarily erased during pericentre passages for eccentric cluster
orbits. By contrast, the concentrated clusters show a
different cluster structure, as they do not seem to show a distinct
core and bulk within the range of the data points.

\subsection{Surface density profiles}\label{Sec:SDP}
\subsubsection{Constant tidal fields}\label{ssec:const}
\begin{figure*}
\includegraphics[width=84mm]{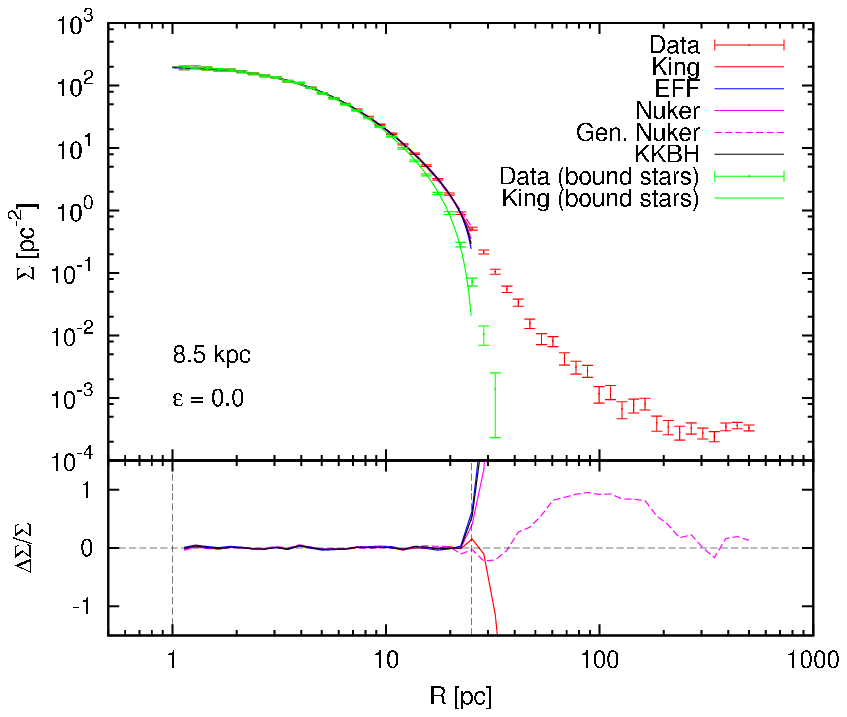}
\includegraphics[width=84mm]{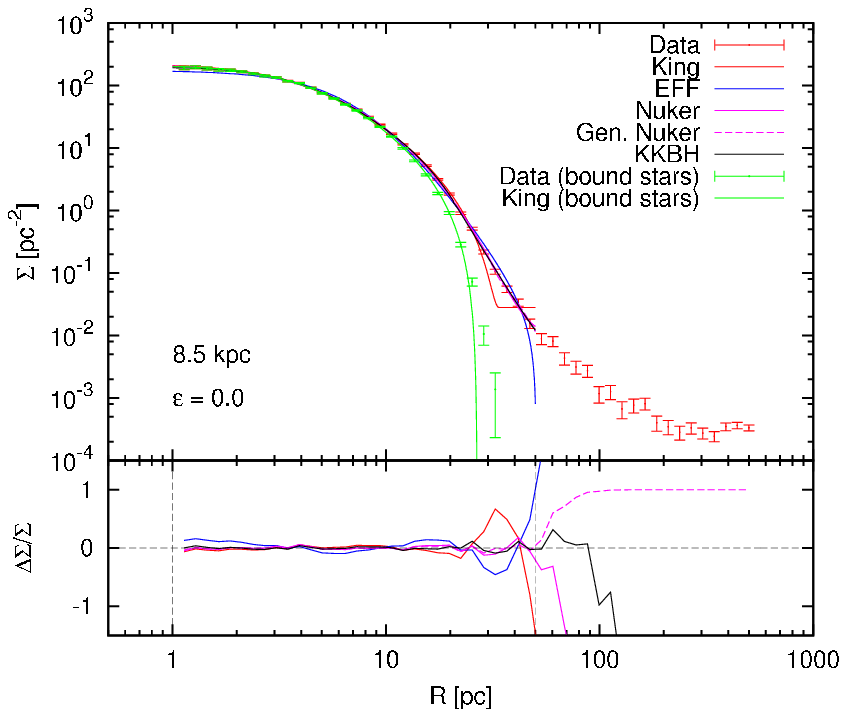}\\
\includegraphics[width=84mm]{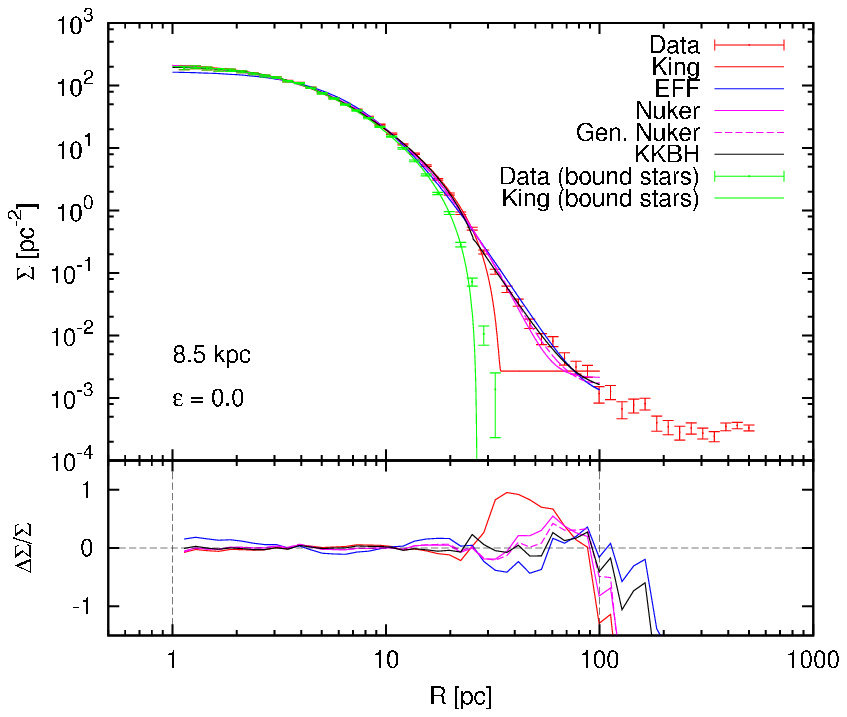}
\includegraphics[width=84mm]{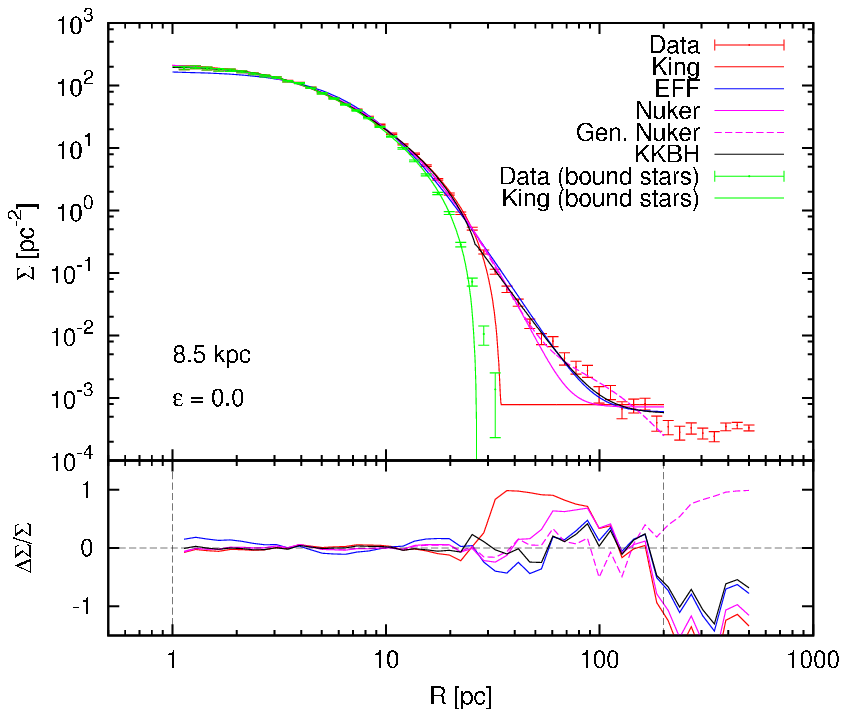}\\
\includegraphics[width=84mm]{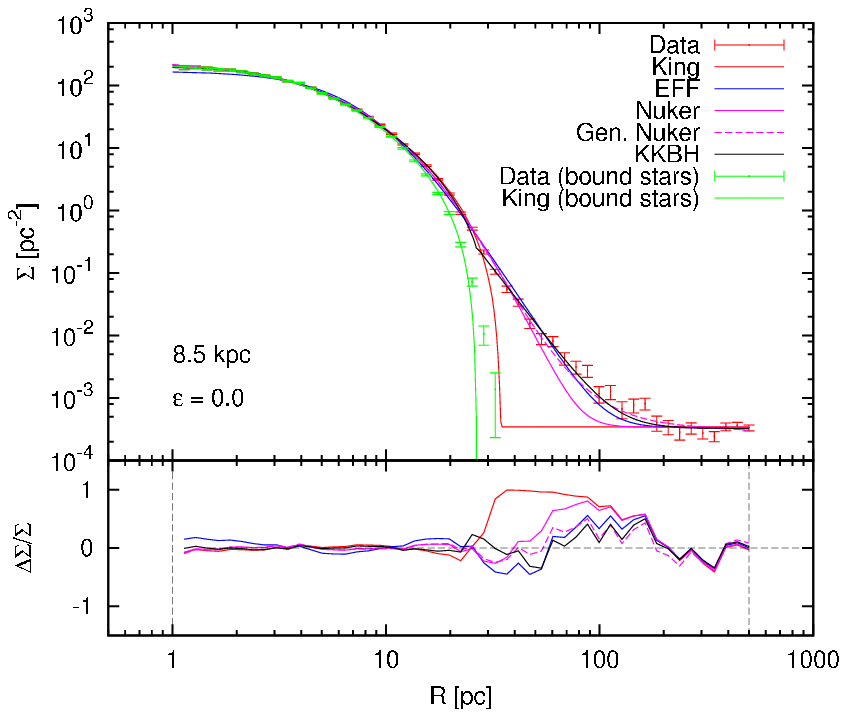}
\includegraphics[width=84mm]{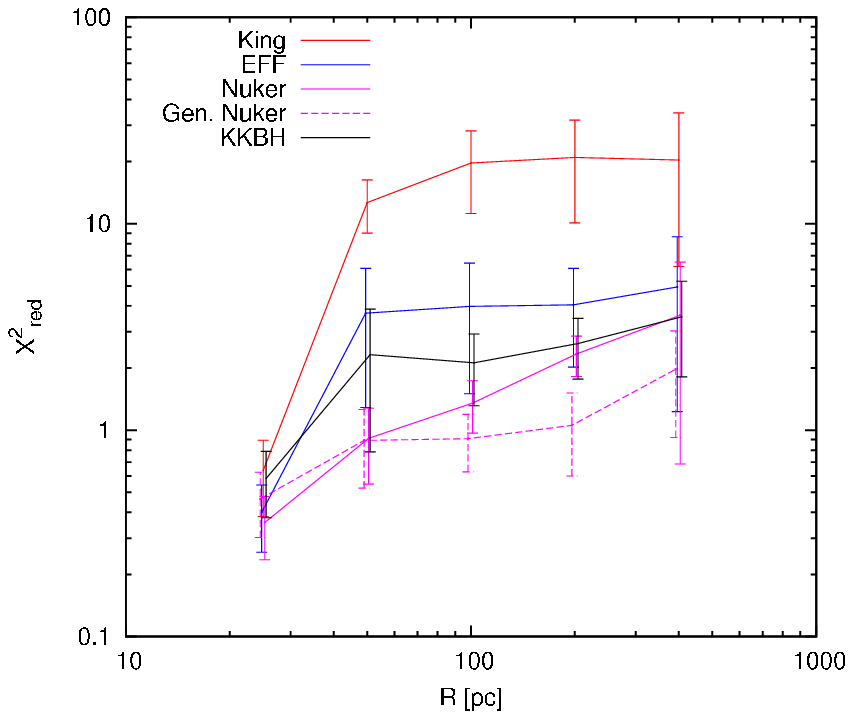}
  \caption{Surface density profiles of the cluster in a constant tidal
  field at a galactocentric distance of 8.5 kpc (B0). The upper panels
  of the first five graphs show the data and the template fits
  according to Sec.~\ref{sec:templates} for the five ranges of 1-[25,
  50, 100, 200, 500] pc. The lower panels of the first five graphs
  give the relative deviation of the templates from the data. Vertical
  dashed lines mark the fitting range. The theoretical Jacobi radius can be read off from the outermost green data point (bound stars). The lower right panel shows the
  reduced $\chi^2$ for all of the five ranges, averaged over all
  snapshots of this cluster. Error bars give the standard deviation of
  $\chi^2_{red}$. A corresponding velocity dispersion profile is given
  in Fig.~\ref{vdps}.}
  \label{sdp8500}
\end{figure*}
\begin{figure}
\includegraphics[width=84mm]{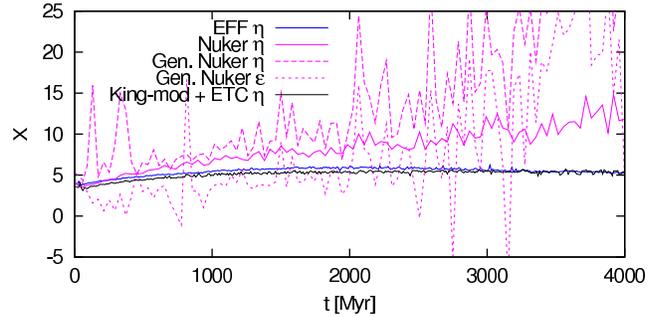}
  \caption{Evolution of the slopes of the tidal debris of the
  templates which can fit such a component, i.e. the surface density
  in the tidal debris falls off with $R^{-X}$, for the cluster on a
  circular orbit at 8.5 kpc (B0). Since the gen. Nuker template has
  two power-law slopes we show both here. Note that the Nuker and the
  gen. Nuker slopes are smoothed, as the variations between the
  individual time steps are very large. The other slopes are not
  smoothed since they are better behaved. The EFF and KKBH slopes show
  a rise in the beginning of the simulation which corresponds to an
  initial increased mass-loss rate due to primordial escapers. The
  Nuker and gen. Nuker slopes give no reliable information on the
  tidal debris slope.}
  \label{eta8500}
\end{figure}
\begin{figure}
\includegraphics[width=84mm]{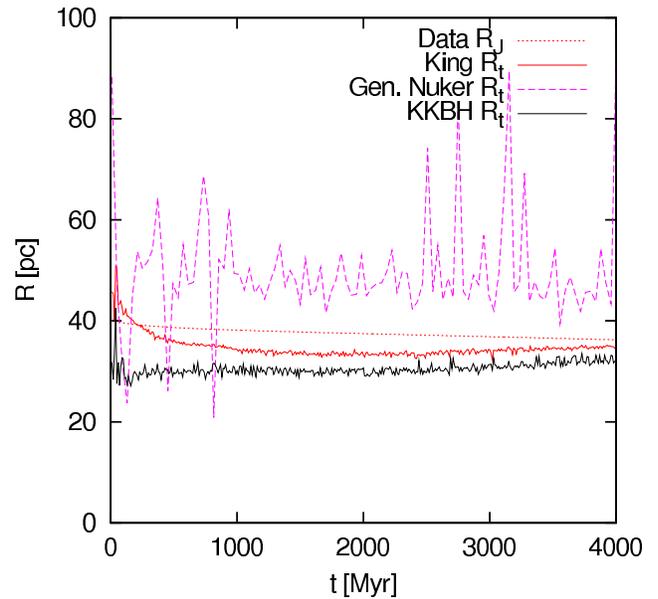}
  \caption{Evolution of the Jacobi radius $R_J$ of the cluster on a
  circular orbit at 8.5 kpc (B0), evaluated using eq.~\ref{eq:R_J}
  (red dotted line), and the edge radii of the fitted templates for a
  fit range of 1-100 pc. None of the edge radii reproduces the Jacobi
  radius well over the whole simulation time. King shows considerable
  evolution with respect to the Jacobi radius. The edge radius of the
  gen. Nuker template is smoothed since it fluctuates a lot, or
  sometimes does not converge properly (see text). In cases when it
  converges, it mostly gives a radius which is about 50\% larger than
  $R_J$. KKBH has about the same values as King, though reduced by
  another 10\%, but is less influenced by the initial mass loss due to
  primordial escapers, i.e. is more stable.}
  \label{rt8500}
\end{figure}
\begin{figure}
\includegraphics[width=84mm]{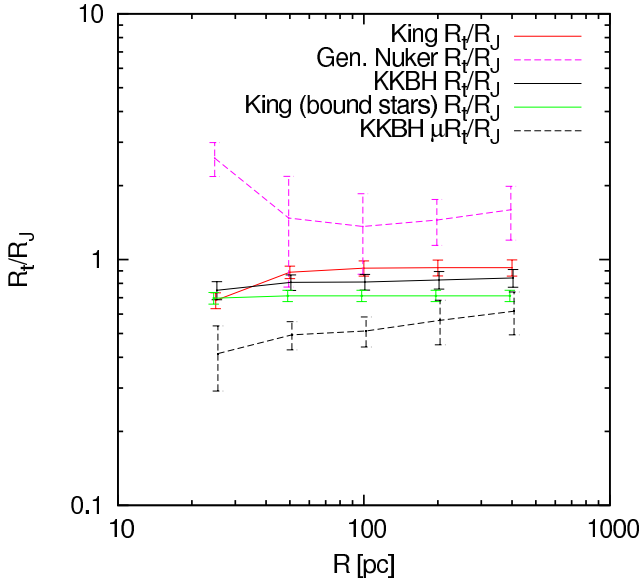}
  \caption{Ratio of the fitted edge radii to the Jacobi radius
  averaged over the whole simulation of the cluster on a circular
  orbit at 8.5 kpc (B0) and shown for all 5 fit ranges. Error bars
  give the standard deviations from the mean. The King template
  reproduces the Jacobi radius to about 10\% accuracy for sufficiently
  large fit ranges. Gen. Nuker gives values 2-3 times larger than the
  Jacobi radius and shows large fluctuations. KKBH shows a constant
  value over all fit ranges which is about 20\% smaller than the
  Jacobi radius. Also shown are the edge radius of the bound stars,
  which lies at about 70\% of $R_J$, and the break radius of the KKBH
  template at which the profile of the cluster becomes dominated by
  potential escapers.}
    \label{r8500}
\end{figure}
As an example to demonstrate the fitting process, in
Fig.~\ref{sdp8500} the surface density profile of the cluster on a
circular orbit at 8.5 kpc is shown when the cluster has a mass of
$15\,000 \msun$, just as was shown for the velocity dispersion in the
4th panel of Fig.~\ref{vdps}. In the upper panels of the first five
graphs the cluster data and the fits of the analytical templates from
Sec.~\ref{sec:templates} are shown for the five different ranges of
1-[25, 50, 100, 200, 500] pc. In the lower panels of these graphs the
relative errors can be seen. The lower right panel gives the reduced
$\chi^2$ for each template for each fit range averaged over all 400
snapshots of the cluster.\\

We see that for the smallest fit range (1-25 pc) all templates fit
about equally well, whereas for the larger ranges significant
differences in the quality of the fits appear. This is due to the fact
that the smallest fit range only covers the core of the cluster, which
in this case is flat, and a part of the bulk. Such a profile can be
fit by all templates in the sample. But already the second smallest
fit range of 1-50 pc covers part of the tidal debris, as the Jacobi
radius in this snapshot is 36 pc. Hence the King template, which does
not have a proper tidal debris term, gets significantly worse with
increasing fit range. Furthermore, since there is an inflexion point
of the slope between the bulk and the tidal debris at about the Jacobi
radius, the templates which cannot account for the bulk and  debris
separately (EFF and Nuker) get worse for increasing fit range as
well.  In contrast  to EFF, however, Nuker can partially compensate the
missing bulk term by a more flexible core term. Besides Nuker, only
the two templates which consist of three parts have a good reduced
$\chi^2$ value over all fit ranges (lower right panel of
Fig.~\ref{sdp8500}).\\

Taking a closer look at the tidal debris of this cluster at the given,
dynamically well-evolved state shows that the slope outside the Jacobi
radius is quite similar to the slope of the bulk, though it is not the same. If
the data had been slightly more noisy we could have even concluded that the
slope of the tidal debris close to the cluster is the same as the
slope of the bulk, which is what \citet{Elson87} might have seen in
their data on young LMC clusters. This is also the reason why EFF and
Nuker fit this cluster reasonably well even though they have no
separate bulk and debris terms.\\

Another  striking fact is  that the tidal debris does not show a
constant slope for radii much larger than the Jacobi radius. From a
constant mass-loss rate (which this cluster shows for several Gyr) and
a constant velocity of the escapers within the tidal tails away from
the cluster we would infer that the number of stars in the debris
grows linearly with radial distance from the cluster. Thus, the
surface density profile of the debris should fall off with $R^{-1}$,
as the surface area of the annuli grows with $R^2$. Instead we see a
slope which is steeper close to the cluster and gets more shallow at
radii of several hundred pc.\\

This is due to the fact that stars do not move linearly along the
tidal tails as they escape from the cluster, but exhibit an epicyclic
motion in which they get periodically accelerated and decelerated
\citep{Kuepper08a}. Since escaping stars are first accelerated away
from the cluster, the slope of the surface density close to the
cluster should be lower than -1. This acceleration then turns into a
deceleration at a distance $R = 0.5\,y_C$, where the length of the
epicycles, $y_C$, depends on the tidal field. At a distance $R = y_C$,
which in this case is given by approximately $3\pi R_J \simeq 340$ pc,
the slowest escapers (escapers with $E-E_{crit} \simeq 0$) are coming
to a halt before they get re-accelerated. Escapers with a finite
velocity above the escape velocity come to a halt at a somewhat larger
$y_C$, so that the mean escaper reaches its lowest drift velocity
along the tidal tails at about 400-500 pc.  (For a detailed description
of the tidal tails of this particular cluster see
\citealt{Kuepper09}). In this range the slope of the surface density
profile should be significantly larger than -1.\\

Of greatest interest with regard to real star clusters is the tidal debris
close to the clusters, as observations barely extend to radii of
several Jacobi radii. Thus, we focus on the inner 100 pc in the
following discussion. In Fig.~\ref{eta8500} the evolution of the
power-law slopes of the tidal debris can be seen as fitted by the
given templates for a fit range of 1-100 pc. We see that EFF and KKBH
have a similar evolution, though KKBH varies less with time. This
is due to the fact that EFF fits one slope for the bulk of the cluster
and the tidal debris, whereas KKBH decouples the two parts and
therefore gives a more accurate value for the slope of the tidal
debris close to the cluster. As expected, the slope is significantly
smaller than -1; indeed it has a value of about -5. Note that KKBH and
EFF show a slightly shallower slope at the beginning of the
simulation, where the cluster loses more mass as a consequence of
primordial escapers (see \citealt{Kuepper09}). By contrast, we
see that the Nuker and the gen. Nuker templates do not give reliable
information on the tidal debris slope. The power-law slopes of the
Nuker and the gen. Nuker templates in Fig.~\ref{eta8500} even had
to be smoothed in order to reduce the stochastic noise to a reasonable
level. Even so, the fluctuations in the values are too large to allow
any conclusions to be drawn from them. This is due to the factors $\alpha$ and
$\delta$ in eq.~\ref{eq:nuker} \& \ref{eq:gennuker} which are meant to
adjust the smoothness of the transition from one slope to another but
have quite a large influence on the fitting results.\\

From Fig.~\ref{sdp8500} we can see that the surface density of bound
stars falls off more steeply than the surface density of all stars in
the sample (green profile vs. red profile), i.e. the edge of the bulk
of bound stars lies further in than the edge of the bulk of all
stars. Thus, the population of potential escapers enhances or even
dominates the surface density profile for radii larger than about half
the Jacobi radius, just as we saw  in the velocity dispersion
profiles (Fig.~\ref{vdps}).   It should be kept in mind, however, that the underlying
investigation is limited to star clusters with up to 64k stars. How
this behaviour changes with increasing $N$ still has to be considered. From
\citet{Baumgardt01}, however, we expect only a weak dependence on $N$.

In Fig.~\ref{rt8500} the time evolution of the Jacobi radius of this
cluster is shown. As the cluster constantly loses mass the Jacobi
radius (eq.~\ref{eq:R_J}) slowly decreases. Also shown in the figure
are the fitted edge radii of the King, the gen. Nuker and the KKBH
templates for a fit range of 1-100 pc. We see that none of the
templates reproduces the Jacobi radius accurately over the whole
simulation time.

The King $R_t$ shows  strong evolution at the beginning, because of  an
initial burst of mass-loss due to primordial escapers, which the template
tries to fit. For the rest of the simulation the King edge radius
agrees with the Jacobi radius to within about 10\%.

The gen. Nuker $R_t$ shows large fluctuations, which are due to the
above mentioned smoothness parameters $\alpha$ and $\delta$,
which introduce a large ambiguity but also cause     problems in the fitting
process. Thus, for this template the Marquardt-Levenberg algorithm
sometimes does not converge properly, and then the starting value for $R_t$,
which was  chosen to be 90 pc for all templates, is retained. In case the
fit converges, it mostly yields an edge radius which is about 50\% too
large.

The KKBH template usually yields similar results to King, though 10\%
smaller, but is much less influenced by the initial mass loss of
primordial escapers, and thus more accurately reproduces the edge radius
of the bulk of the cluster. This case is a good example in which heavy
mass loss significantly influences fitting results, as may have
happened in the investigation of \citet{Elson87}, or as was mentioned
by \citet{Cote02}, \citet{Carraro07} and \citet{Carraro09} for the
Milky-Way globular clusters Palomar 13, Whiting 1 and AM 4,
respectively. The structure of the KKBH template helps to minimise
such a problem.\\

In Fig.~\ref{r8500} we show all these fitted edge radii in units of
the Jacobi radius, averaged over all time steps of the simulation, for
all 5 fit ranges. King shows a value which lies within 10\% of the
true Jacobi radius for large fit ranges. Only the smallest fit range
of 1-25 pc yields an average edge radius which is about 30\% off, so
here the fit range would significantly influence the fitting
result. In contrast to that the KKBH template gives a more constant
value over all fit ranges but which is reduced compared to King by
about 10\%. The gen. Nuker template does not yield a reliable value
for the edge radius, as the fluctuations between the time steps are
large, as indicated by the error bars, and the mean strongly depends
on the fit range.\\

Throughout the analysis we found the bound stars to be well
represented by a simple King template. From Fig.~\ref{r8500} we see
that the edge of bound stars lies at about 70\% of the Jacobi
radius. The break radius of the KKBH template, at which it changes
into the power-law tidal debris, is also shown in the figure and lies
at about 50\% of $R_J$ and can be interpreted as the radius at which
the profile of the cluster deviates from a smooth bulk profile and
gets dominated by potential escapers.\\

In Tab.~\ref{table3} a summary of the fitting results of the edge
radii for the three smallest fitting ranges for all clusters in a
constant tidal field is given. From this table we can conclude that
the edge radii are closer to the Jacobi radius the larger the fit
range is. For the smallest fit range the deviation can be even up to
40\%. Averaging over all the fitting results in Tab.~\ref{table3}
confirms our findings: the King edge radius is on average a little
(10-20\%) smaller than the true Jacobi radius. Gen. Nuker yields no
reliable results, whereas KKBH is more stable than King but yields an
edge radius which is also about 20\% smaller than the true Jacobi
radius. Furthermore we see from the table that the edge of bound stars
lies at about 70\% of the Jacobi radius and that KKBH finds a break in
the profile slope at about 50\% of the Jacobi radius. Thus, at this
radius the profile becomes dominated by potential escapers.\\

From the investigation of the clusters in a constant tidal field we
can conclude that our ansatz for constructing a star cluster template
consisting of core, bulk and debris was chosen well. The unbound tidal
debris has a significant influence on the fitting results, but this can be
minimised by the addition of an independent tidal debris term. In
addition, our constructed KKBH template is a good tool with which to measure the
power-law slope of the tidal debris and to look for a break in the
surface density profile.  In this way we saw that the influence of potential
escapers on the cluster profile is significant for radii larger than
0.5$R_J$ and that beyond about 0.7$R_J$ nearly all stars are unbound.

\subsubsection{Time-dependent tidal fields}\label{ssec:timedep}
\begin{figure*}
\includegraphics[width=84mm]{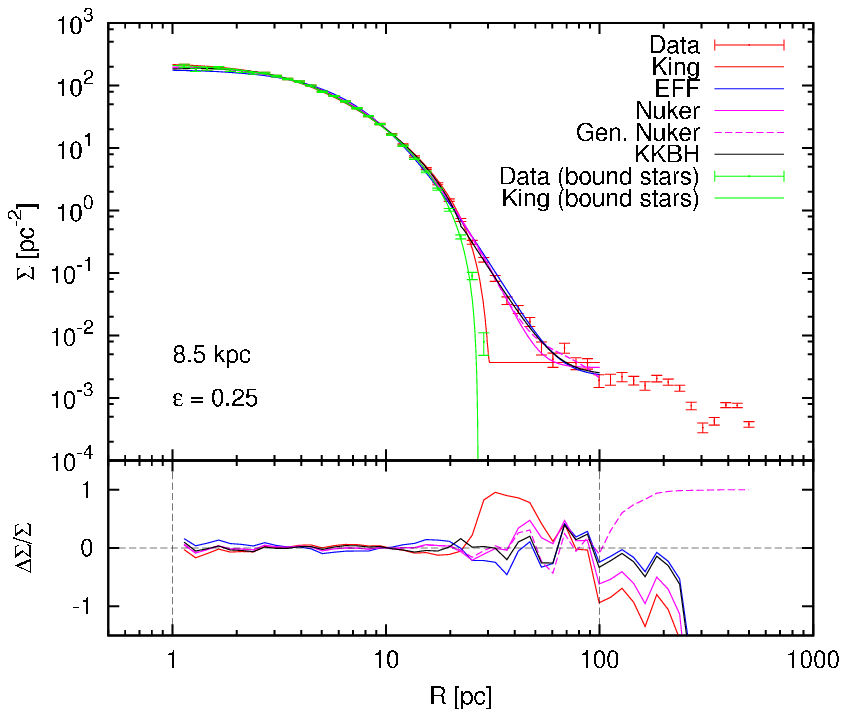}\hspace*{3mm}
\includegraphics[width=84mm]{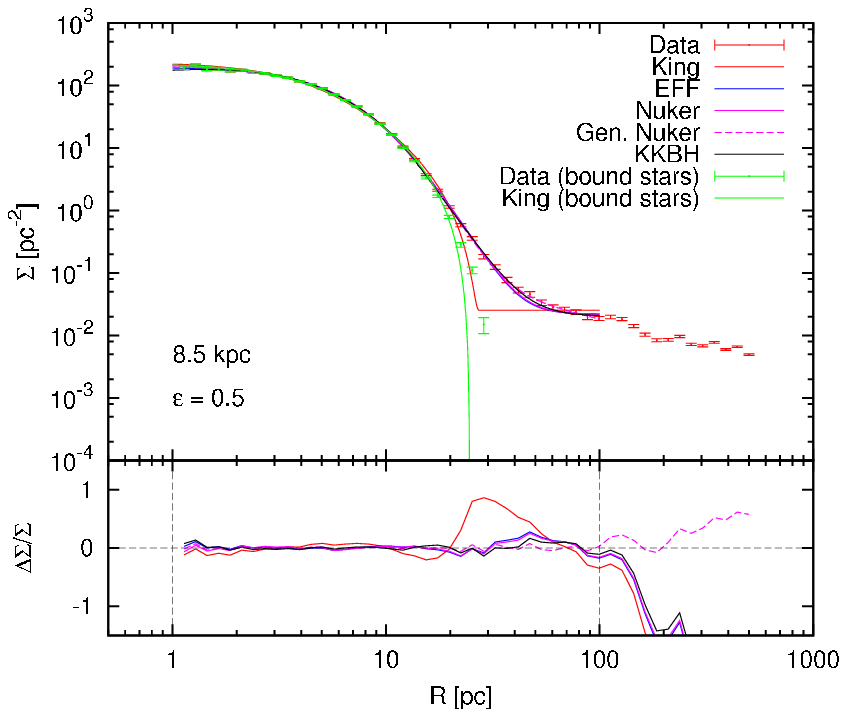}\\
  \caption{Surface density profiles of the clusters at an apogalactic
  distance of 8.5 kpc with an orbital eccentricity of 0.25 (B1, left)
  and 0.5 (B2, right).  At the time of the snapshot both clusters are 
  at apogalacticon and have a mass of about $15\,000\msun$. The
  power-law slope of the tidal debris as measured by the KKBH template
  is $\eta = 5.2\pm0.3$ on the left and $\eta = 4.5\pm0.2$ on the
  right. The inner cluster profiles look quite similar whereas the
  average density outside the cluster with an orbital eccentricity of
  0.5 is higher, which is due to the higher mass-loss rate of this
  cluster and the larger orbital compression of its tidal tails. The theoretical Jacobi radii can be read off from the outermost green data points (bound stars). Corresponding velocity dispersion profiles are given in
  Fig.~\ref{vdps}.}
  \label{sdp8500ecc}
\end{figure*}
\begin{figure}
\includegraphics[width=84mm]{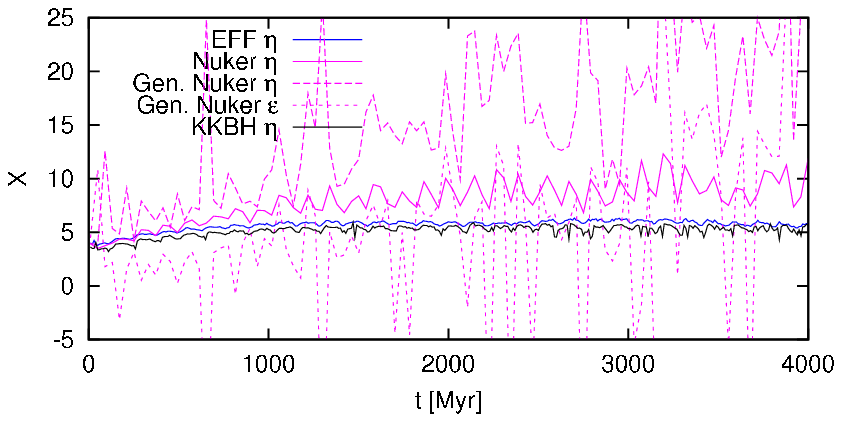}\\
\includegraphics[width=84mm]{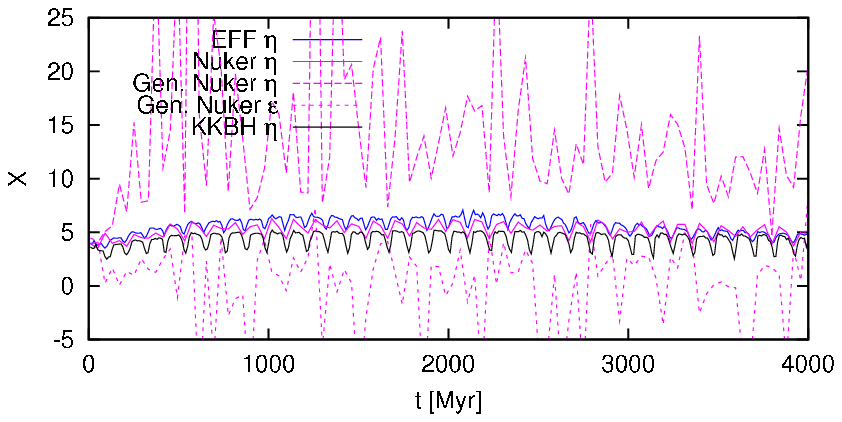}
  \caption{Evolution of the slopes of the tidal debris of the
  templates as shown in Fig.~\ref{eta8500} for the clusters with an
  apogalactic distance of 8.5 kpc and an orbital eccentricity of 0.25
  (B1, upper panel) and 0.5 (B2, lower panel). The Nuker and the
  gen. Nuker slopes are smoothed with a cubic spline, as the
  variations between the individual time steps are very large. The EFF
  and KKBH (for $\epsilon = 0.5$ also the Nuker) slopes again show a
  similar evolution around an average value of about $\eta = 5$. For
  increasing orbital eccentricity the power-law slope of the tidal
  debris starts to vary with the orbital phase. As the cluster gets
  compressed close to apogalacticon, the slope of the tidal debris
  abruptly decreases to values of about $\eta = 2-3$.}
  \label{eta8500ecc}
\end{figure}
\begin{figure*}
\includegraphics[width=54mm]{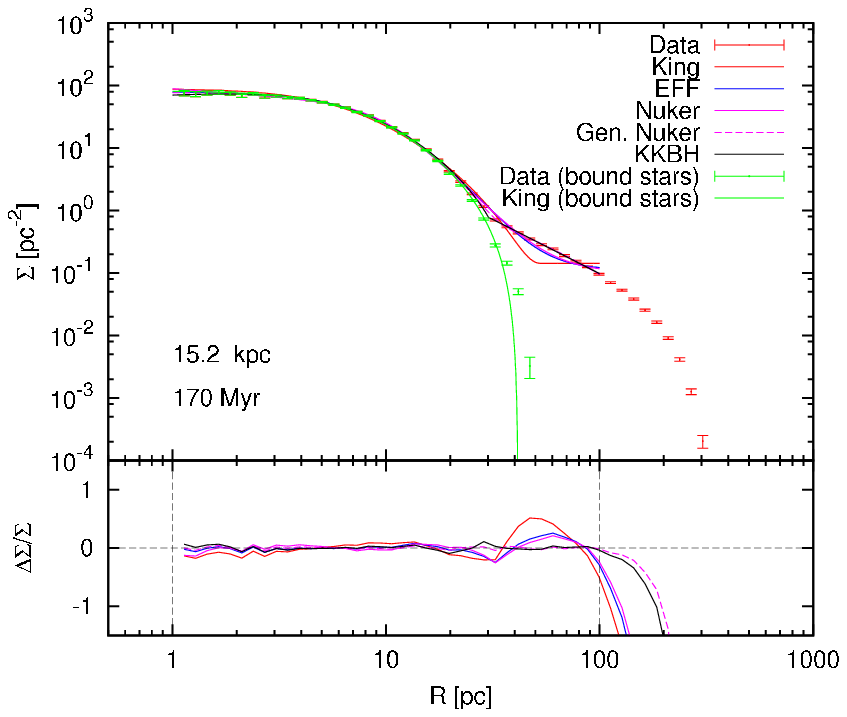}
\includegraphics[width=54mm]{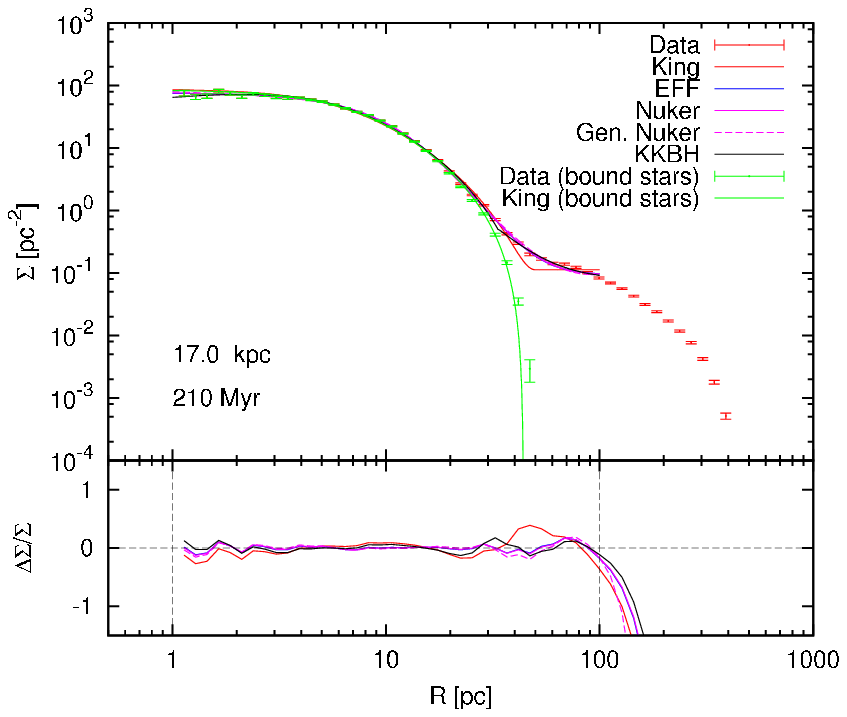}
\includegraphics[width=54mm]{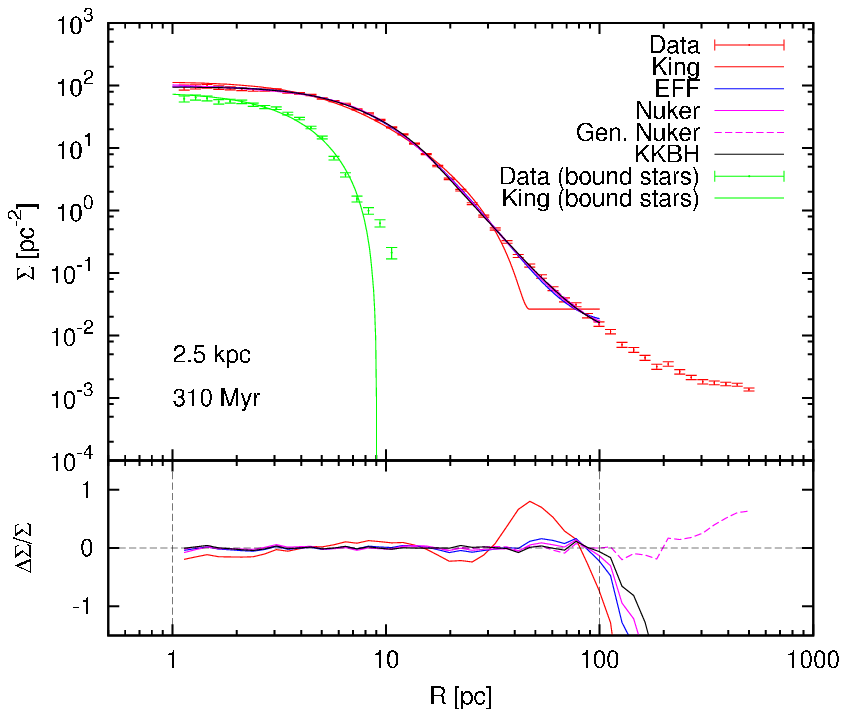}\\
  \caption{Surface density profiles for the cluster at an apogalactic
  distance of 17 kpc with an orbital eccentricity of 0.75 (D3). On the
  left the cluster is shown shortly  before reaching apogalacticon, in the
  middle it is at apogalacticon whereas on the right it is at
  perigalacticon. The power-law slope of the tidal debris as measured
  by the KKBH template is $\eta = 1.7\pm0.2$ on the left, $\eta =
  2.9\pm0.5$ in the middle and $\eta = 3.8\pm0.1$ on the right. Close
  to and at apogalacticon the cluster and its tidal tails get heavily
  compressed which leads to a great enhancement in the number of  stars close to
  the cluster.  At perigalacticon most of the profile is best fit by a
  power-law. Furthermore, most stars are unbound (green line), just as
  we have seen in the velocity dispersion profile (middle graph of
  Fig.~\ref{vdpse075}). The drop in surface density in the left and
  middle panel for large radii is due to the young age of the cluster
  and its short tails, as escaped stars simply could not travel  far at this time.
  Corresponding velocity dispersion profiles are given in
  Fig.~\ref{vdpse075}.}
  \label{sdp17000}
\end{figure*}
\begin{figure}
\includegraphics[width=84mm]{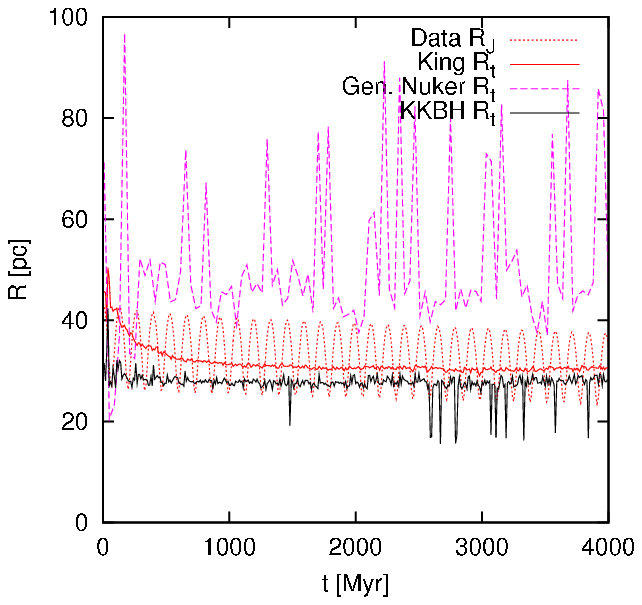}\\
\includegraphics[width=84mm]{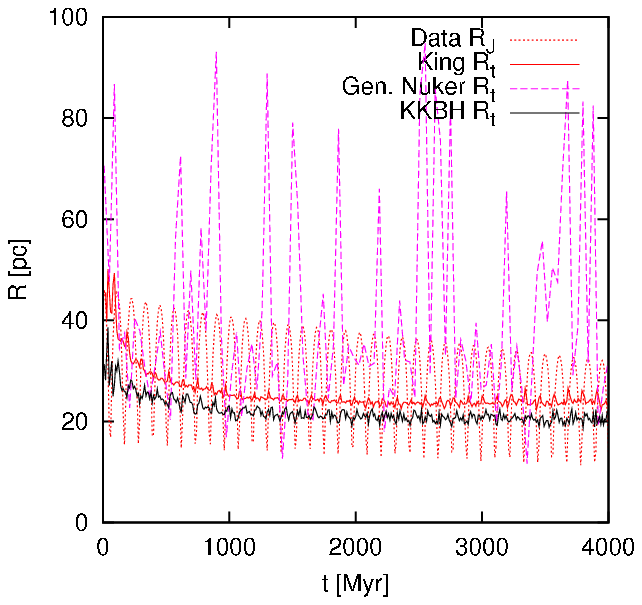}
  \caption{Evolution of the Jacobi radii, evaluated using
  eq.~\ref{eq:R_J} (red dotted line), and the edge radii of the fitted
  templates for a fit range of 1-100 pc for the clusters with an
  apogalactic distance of 8.5 kpc and an orbital eccentricity of 0.25
  (B1, upper panel) and 0.5 (B2, lower panel). The line colours are
  the same as in Fig.~\ref{rt8500}. Even though the Jacobi radius
  changes periodically the bulk of the cluster seems to be
  unaffected. The evolution of the King and KKBH edge radii (red and
  black solid line) are very much like in Fig.~\ref{rt8500} but
  shifted to lower values for increasing orbital eccentricity. Only
  the break radius of the KKBH template (black dashed line) seems to
  be somewhat affected by the changing tidal field.}
    \label{rt8500ecc}
\end{figure}
In Fig.~\ref{sdp8500ecc} we see the clusters with an orbital
eccentricity of 0.25 (left panel) and 0.5 (right panel) at the time
when they are at their apogalactic distance of 8.5 kpc, just as shown
in Fig.~\ref{vdps} for the velocity dispersion profiles. Also shown in
the figure are the template fits for a fit range of 1-100 pc. We see
no significant difference from the surface density profile of the
cluster on a circular orbit (Fig.~\ref{sdp8500}), except for the mean
density within the tidal debris, which gets larger with increasing
eccentricity. This is due to the average mass loss of the clusters as
well as the orbital compression of the tails at apogalacticon, which
both increase with eccentricity.\\

The slope of the tidal debris close to the cluster ($\sim R^{-\eta}$)
is  $\eta = 5.2 \pm 0.3$ for $\epsilon = 0.25$ and $\eta = 4.5 \pm
0.2$ for $\epsilon = 0.5$, i.e. about the same as in the
constant tidal field (about $\eta = 5$ for $\epsilon = 0.0$). But
since the cluster has a lower orbital velocity  at apogalacticon for
increasing eccentricity, the epicyclic overdensities in the tidal
tails are also located closer to the cluster (see \citealt{Kuepper09})
and the surface density is enhanced relative  to the circular case.\\

From Fig.~\ref{eta8500ecc} we see that for a small orbital
eccentricity of $\epsilon = 0.25$ the tidal debris slope evolves very
much as in the constant tidal field case. For $\epsilon = 0.5$ we
see that the slope on average shows the same evolution, but in
addition shows jumps to smaller values. These jumps occur shortly
before apogalacticon where the tails get maximally compressed due to
the orbital acceleration, so that, for large eccentricities, they
literally get pushed back inside the cluster. This enhances the tidal
debris close to the cluster and thus temporarily yields a shallower
slope. This effect is shown in Fig.~\ref{sdp17000} for the cluster
with an apogalactic distance of 17 kpc and an orbital eccentricity of
0.75. The left panel shows the cluster close to apogalacticon, the
middle panel shows it at apogalacticon whereas the right panel shows
the cluster at perigalacticon, similar to what we have seen in
Fig.~\ref{vdpse075}. The slopes $\eta$ of the three snapshots as
measured by the KKBH template are $\eta = 1.7\pm0.2$ close to
apogalacticon, $\eta = 2.9\pm0.5$ at apogalacticon and $\eta =
3.8\pm0.1$ at perigalacticon.\\

Summarising the results, including also those from all the other models of our systematic
study, which are not shown here in detail, we can conclude that the
tidal debris close to the cluster (up to about three times the Jacobi
radius) has a power-law slope $\eta$ in the range 4 to 5, which can decrease to
values in the range 1-2 shortly before apogalacticon due to orbital
compression of the tidal tails. This may lead to the conclusion that
we can identify those star clusters which are at an orbital phase close to
apogalacticon by the slope of their surface density profiles outside
the Jacobi radius. Star clusters like Palomar 13, AM 4 and Whiting 1
may well be in such an orbital phase, as their surface density
profiles show slopes of about $\eta = 1.8$ \citep{Cote02, Carraro07,
Carraro09}. This assumption is further strengthened by the
Milky-Way globular cluster Palomar 5, whose orbit is well constrained
due to its long tidal tails \citep{Dehnen04}. It is currently located
close to apogalacticon and shows a power-law slope of about -1.5
\citep{Odenkirchen03}.\\

From the right panel of Fig.~\ref{sdp17000} we can see that at
perigalacticon, where the Jacobi radius is smallest (about 10 pc),
most of the cluster is unbound (green data points vs. red data points)
just as we have seen in the corresponding velocity dispersion profile
(middle panel of Fig.~\ref{vdpse075}). Also note here how much more poorly
the King template fits the cluster at perigalacticon than at
apogalacticon. By contrast, the KKBH template, as well as the others
 with a power-law tidal debris term, can cover the whole
cluster and its debris perfectly well.\\

In Fig.~\ref{rt8500ecc} we show the evolution of the Jacobi radius and of
the fitted edge radii for the above clusters with eccentricities of
0.25 and 0.5, just as in Fig.~\ref{rt8500} for $\epsilon = 0.0$. While
the Jacobi radius  oscillates with an amplitude which increases with increasing
eccentricity, the overall evolution of the other radii is very similar
to the constant tidal field case. The edge radius of the King template
shows some evolution at the beginning but then gives the mean Jacobi
radius of the cluster. The KKBH template again shows less evolution
than King at the beginning, i.e. is less affected by heavy initial mass loss,
but again gives a nearly constant value similar to that of King, but reduced by
about 10\%. The gen. Nuker edge radius does not show a well-behaved
evolution at all.\\

The results from all models on eccentric orbits in our set are
presented in Tab.~\ref{table4a}-\ref{table4c} sorted by
eccentricity. The results of the clusters with $\epsilon = 0.25$ and
$\epsilon = 0.5$ agree well with the findings for the clusters in a
constant tidal field (Tab.~\ref{table3}).  We note especially the consistent
values for the edge of the
bound stars at 70\% of the Jacobi radius and the break radius of the
KKBH template at 50\% of the Jacobi radius. But also
the fact that the King template yields on average about 90\% of the
\textit{mean} Jacobi radius is interesting, since it is commonly
thought that observed edge radii correspond to the
\textit{perigalactic} Jacobi radii of the corresponding clusters. The fact that
the edge radius is not set at perigalacticon
is due to the fact that stars do not leave the cluster immediately
after becoming unbound but stay within the Jacobi radius for up to
several orbits of the cluster about the galaxy. Moreover, when we
consider the elongated shape of the equipotential surfaces of a
cluster in a tidal field it is quite reasonable that a spherically
symmetric, one-dimensional density profile will yield a smaller value
than the two-dimensional (i.e. projected) Jacobi radius.\\

Note that the mean results and fluctuations get worse for increasing
tidal field strength. All clusters with orbital eccentricities of 0.75
get disrupted within a few orbits about the galactic centre and thus
are barely in a state close to virial equilibrium. Also the clusters
at 4.25 kpc and 8.5 kpc with eccentricities of 0.5 show large spreads
in the fits as they penetrate deeply into the central part of the galaxy. At the
beginning of these $N$-body computations, however, when the clusters are still
very massive and much less vulnerable to tidal influences, the fitting
results are more stable and agree with the findings for the other
clusters.\\

From the clusters in time-dependent tidal fields we can conclude that
our KKBH template again proves to be a good tool to investigate the
state of the clusters and their tidal debris. Thus, we were able to
extend our findings of the previous section to clusters on eccentric
orbits. Moreover, we suggest that star clusters in an orbital phase
close to apogalacticon can be identified by the slope of their surface
density profile outside the Jacobi radius. We furthermore find that
the bulk of the cluster adjusts to the mean tidal field rather than,
as usually believed, to the perigalactic tidal field.

\subsubsection{Concentrated clusters}\label{ssec:conc}
\begin{figure*}
\includegraphics[width=84mm]{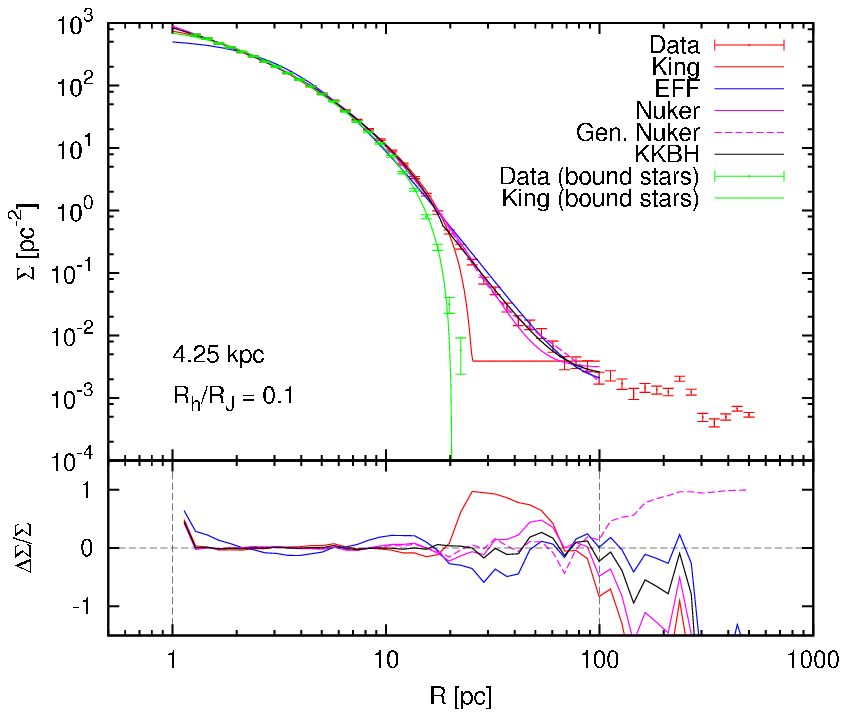}
\includegraphics[width=84mm]{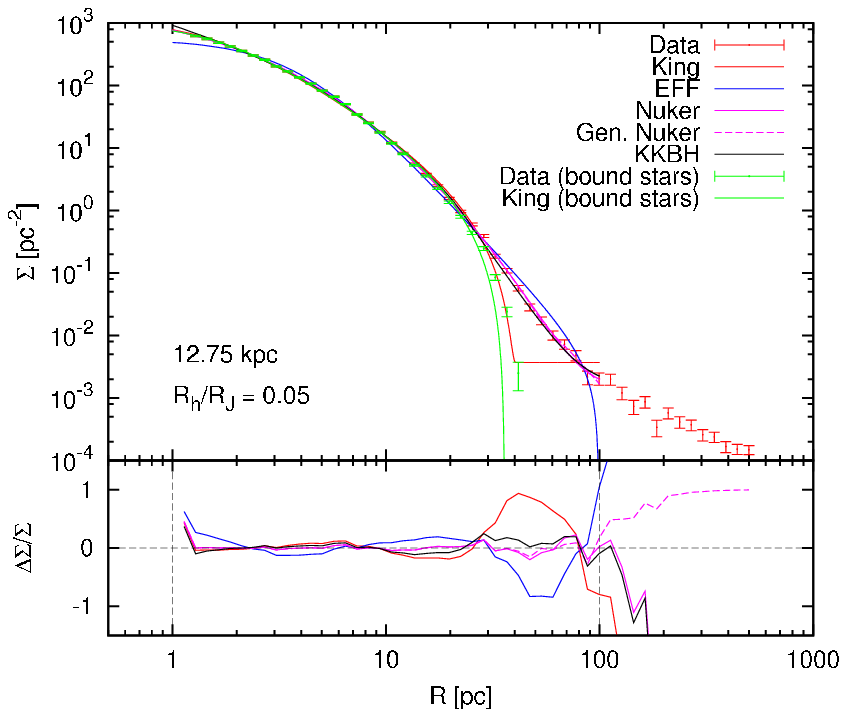}
  \caption{Surface density profiles of the concentrated clusters at a
  galactocentric radius of 4.25 kpc (A0c, left) and 12.75 kpc (C0c,
  right).  At the time of the snapshots both clusters have  a mass of
  about $15\,000\msun$. The power-law slope of the tidal debris as
  measured by the KKBH template is $\eta = 4.6\pm0.6$ (left) and $\eta
  = 4.3\pm0.3$ (right). The theoretical Jacobi radii can be read off from the outermost green data points (bound stars). One corresponding velocity dispersion profile is  given in Fig.~\ref{vdpcc}.}
  \label{sdpcc}
\end{figure*}
Fig.~\ref{sdpcc} shows the surface density profiles of the clusters at
4.25 kpc (left) and 12.75 kpc (right), which both had an
initial (3-dimensional) half-mass radius of 3 pc; for the first of these the
corresponding velocity dispersion profile was shown in
Fig.~\ref{vdpcc}. At the time of the snapshots the clusters have
$15\,000\msun$ and a ratio of projected half-mass radius to Jacobi
radius of 0.08 (left) and 0.06 (right). From the figures and the
fitted templates we see that the power-law slope of the tidal debris
agrees with the previous findings ($\eta = 4.6\pm0.6$ for the left
panel and $\eta = 4.3\pm0.3$ for the right). Furthermore, we see that
the EFF template completely fails to fit the data. This is due to the
fact that the clusters do not have a distinct core (at least down to a
radius of 1 pc) but appear to consist only of a bulk and tidal
debris. Thus the EFF template, which consists of a flat core and a
tidal debris term, cannot reproduce it. The King template does a better
job for the bulk but fails at the edge of the bulk, whereas  KKBH
can give a good representation of all parts of both clusters. Nuker and
gen. Nuker are also flexible enough to cover the whole radial
range. These characteristics become more pronounced for increasing
galactocentric radius as the half-mass radius is fixed at 3 pc but the
Jacobi radius increases, and so the concentration increases. Note also
that the quality of the King template fit decreases with a larger
concentration. This supports the findings by \citet{Baumgardt10} that
concentrated clusters cannot be properly fitted by a King template.\\

But even though most templates yield good $\chi^2$ values, from
Tab.~\ref{table5} we can see that the fitted edge radii do not
correspond to the Jacobi radii any more. With increasing concentration
the edge radii shrink compared to the Jacobi radii. That is the reason
why we avoid the term tidal radius and instead use edge radius, since
this edge radius does not have to be a result of the galactic tide.\\

Moreover, since the bound stars follow a similarly concentrated
distribution and since we tried fitting this distribution with a King
template, these numbers also get unreliable. Here a reconsideration of
the templates with regard to the structure of concentrated clusters
would be necessary, since the templates we used here all seem to fail
in giving information on the cluster structure, due to the fact that
these clusters do not show a distinct core and bulk down to 1
pc. Thus, we conclude that Jacobi radii cannot be reliably determined
from observed clusters using these templates if their concentration is
high, i.e. with $R_{hp}/R_J \simeq 0.1$ or smaller.

\subsection{Review of templates}
Before summarising the results of this investigation let us briefly
discuss the performance of the templates which were defined in
Sec.~\ref{sec:templates}  and which were used here.
\begin{itemize}
\item \textbf{King} has proved to be a reasonable tool for studying
the Jacobi radius of extended clusters but fails when the cluster has
a pronounced tidal debris component or is too concentrated. It  cannot give
information on the core slope of either mass segregated clusters or clusters
with an IMBH in the centre. Eccentric cluster orbits are also a
problem for the template. Moreover, the fit results strongly depend on
the  radial range of the cluster which is covered; it is thus quite unreliable
for obtaining cluster parameters.
\item \textbf{EFF} is a good tool for the study of clusters with a pronounced
tidal debris component, provided that they  are not too concentrated or mass segregated. It
cannot, however, give information on the edge radius.
\item \textbf{Nuker} can fit most of the clusters in our sample well,
but it is not possible  to deduce cluster parameters  reliably from
the fitting results.  Furthermore it does not have an edge radius.
\item \textbf{Gen. Nuker} is a very powerful template with a low
$\chi^2$ for all fit ranges but, just as Nuker, does not allow 
the deduction of reliable structural parameters from the fitting results, as
the resulting parameters of the template vary strongly and depend
strongly on the radial coverage of the cluster.
\item \textbf{KKBH} has proved to be a reasonable extension of King,
improving the fitting results for all clusters in the sample. It also
yields the most reliable structural information over all fit ranges in
terms of edge radius and the slope of the tidal debris. Like all other templates, however,
it fails in the case of concentrated clusters.
\end{itemize}

\section{Conclusions}\label{Sec:Conclusions}
This systematic investigation of velocity dispersion profiles and
surface density profiles of star clusters with masses of about $10^4
\msun$ has shown how complex and versatile these profiles can be,
depending on the circumstances of the clusters.\\

Furthermore we have shown the importance of potential escapers for
these two quantities. Both profiles are influenced by this population
of energetically unbound stars for radii larger than about 50\% of the
Jacobi radius and are completely dominated by it for radii larger than
about 70\% of the Jacobi radius (Fig.~\ref{vdps} \&
\ref{vdpse075}). \citet{Baumgardt01} found that for clusters in a
constant tidal field the fraction of potential escapers varies with
the number of stars within a cluster approximately as
$N^{-1/4}$. Thus, even if the cluster mass increases by a factor of
10, the fraction (and influence) of potential escapers will only be reduced
by less than 50\% compared to this investigation.\\

As a consequence of the potential escaper population, the velocity
dispersion profiles in our sample do not show a clear separation
between the bulk of the cluster and the tidal debris at the Jacobi
radius, but rather show a smooth transition from cluster to
debris. For mass estimates which are based on velocity dispersion
measurements we therefore recommend to use stars within 50\% of the
Jacobi radius to minimise the effect of potential escapers.\\

Moreover, we argue that investigations on velocity dispersion profiles like \citet{Drukier98}, \citet{Scarpa03} or \citep{Scarpa07} have to be interpreted with caution. Also, detecting a possible dark matter halo around globular clusters is less feasible due to the population of potential escapers. Based on our investigation we suggest that no deviation from Newtonian gravity in a star cluster has been detected so far, neither has there been evidence for a dark matter halo.\\

From our large set of surface density profiles we saw that in most
cases the structure of extended star clusters can be split into core,
bulk and tidal debris.
\begin{enumerate}
\item The core is the innermost part of the cluster which extends out
to the core radius.  In most cases it is flat, and only for clusters with
strong mass segregation or with an IMBH is it cuspy. Very concentrated
clusters with $R_{hp}/R_J < 0.1$ have such a small core that fitting
such clusters leads to problems with templates as well as with
physical models (see also \citealt{Baumgardt10}).\\\\
\item The bulk contains most stars of the cluster and extends from the
core to the tidal debris. We found that the bulk can be well
represented by a King-like template. We named the radius at which such
a template approaches zero density the edge of the bulk and not the tidal
radius of the cluster, since we found it not to be consistent with the
Jacobi radius. In the case of concentrated clusters, the edge radius
is not even determined by tidal forces  at all (Tab.~\ref{table5}).\\

Moreover, for clusters on eccentric orbits we found that the edge
radius of a cluster adjusts to the mean Jacobi radius
(Fig.~\ref{rt8500ecc}) and not the perigalactic Jacobi radius, which has been
assumed in earlier investigations (e.g. \citealt{Innanen83},
\citealt{Fall01}), but has never been checked by self-consistent
calculations on a star-by-star basis. Since nearly all globular
clusters of the Milky Way are more compact and massive than our test
cluster, and thus should be less influenced by tidal variations, this
finding should also hold for most of these clusters.\\

The edge radius of extended and moderately concentrated clusters, as
fitted by the King template, lies within about 10\% accuracy at ~90\% of
the mean Jacobi radius (Tab.~\ref{table3}-\ref{table4c}). But we found
the King template to be a bad representation of the true profile in
the case of non-circular cluster orbits because it is significantly
influenced by the tidal debris and by the potential escaper
population. Furthermore the accuracy of the results strongly depends
on the radial coverage of the cluster. We therefore created an
enhanced version of the King template which has more flexibility in
the core and has an additional tidal debris term, given by a
power-law. With this KKBH template we achieved a higher stability over
all fit ranges. Furthermore, we were able to measure the edge of the
bulk more accurately and found it to lie at 80\% of the Jacobi
radius. This discrepancy is easily understandable if we take into
account the fact that the edge radius of a template fits the azimuthally
averaged mean of a cluster's tidal surface, whereas the Jacobi radius
is by definition the semi-major axis of the equipotential surface.\\

Furthermore we found that the edge radii from concentrated clusters
with $R_{hp}/R_J < 0.1$ cannot be extracted properly with currently
available templates. The higher the concentration of the cluster the
smaller is the ratio of the fitted edge radius to the theoretically evaluated
Jacobi radius (Tab.~\ref{table5}).\\\\

\item The tidal debris falls off like a power-law with a slope of about -4 to
  -5, rather than the  expected
dependence of  $R^{-1}$. This is due to
the epicyclic motion of the stars in the tidal tails, which is most
pronounced in the vicinity of the cluster.\\

For clusters on eccentric orbits, at a time shortly before reaching
apogalacticon, the slope was found to deviate for a short time to a
value of about -2 as a consequence of orbital compression of the tails
(Fig.~\ref{eta8500ecc} \& \ref{sdp17000}). We suggest that this
enhanced slope can be used to identify star clusters which are close
to or at the apogalacticon of their orbit. The most prominent example of
such a cluster is the Milky Way globular cluster Palomar 5 which is
well known to be close to the apogalacticon of its orbit \citep{Dehnen04},
and which shows a power-law slope of -1.5 outside its assumed Jacobi
radius \citep{Odenkirchen03}. Moreover, the MW globular clusters
Palomar 13, AM 4 and Whiting 1 may be further candidates for being
close to apogalacticon, since their surface density profiles show
power-law slopes of about -1.8 outside the Jacobi radius, which has been 
attributed to heavy ongoing mass loss by the authors of the
corresponding investigations \citep{Cote02, Carraro07, Carraro09}.\\

Furthermore we showed that due to the population of potential escapers
the power-law slope of the tidal debris  begins even before  the
Jacobi radius is reached. Thus, the bulk and the tidal debris partially
overlap. With our KKBH template we found the debris to dominate for
radii larger than about 50\% of the Jacobi radius
(Tab.~\ref{table3}-\ref{table4c}).
\end{enumerate}

With the help of our comprehensive set of clusters it is possible to
tailor a star cluster template to the  facts we have found. Our KKBH template
seems to be a good first step but, like all other available templates,
shows discrepancies for concentrated clusters. Of course, templates
are less attractive than physically motivated models. None of the
existing models, like \citet{King66} and \citet{Wilson75}, however, account for
the above mentioned facts.  We suggest that a general,
physically motivated model for star clusters should include a term for
mass segregation and should include a potential escaper population when
being fitted to observations. Finally, the possible influence of
(compressed) tidal tails on the surface density profiles of star
clusters should be kept in mind when fitting surface density
profiles.\\

Moreover, the fact that all clusters in the sample showed a constant
edge of the bulk with time (Fig.~\ref{rt8500ecc}) and that stars which
get unbound escape from a cluster with a delay, supports our
theoretical treatment of epicyclic overdensities in tidal tails for
clusters in time-dependent tidal fields in \citet{Kuepper09}.

\section*{Acknowledgements}
We are very grateful to Sverre Aarseth for making his \textsc{Nbody4} and \textsc{Nbody6} codes freely vailable and for continuous support. AHWK kindly acknowledges the support of an ESO Studentship.

\appendix
\section{Fitting KKBH using GNUPLOT}
When using \textsc{gnuplot}, fitting the KKBH template is quite simple
since it is possible to use ternary operators. First  the two
functions $f_1(R)$ and $f_2(R)$ (eq. \ref{eq:ETC1} \& \ref{eq:ETC2}) need to be defined with\\\\ \texttt{f1(x) = (x<Rt ? k * (x/Rc/(1.0+x/Rc))**(-gamma) *
(1.0/(1.0+(x/Rc)**2.0)**(0.5) - 1.0/(1.0+(Rt/Rc)**2.0)**(0.5))**2.0 :
0)}\\\\ and\\\\ \texttt{f2(x) = f1(Rt*mu) * (1.0 +
(x/(Rt*mu))**64.0)**(-eta/64.0)}\\\\ after which  the KKBH
function is defined\\\\ \texttt{f(x) = (x < mu*Rt ? f1(x) + b : f2(x) + b)}\\\\
is defined.
Then the constants  have to be set to some reasonable value, e.g.\\\\
\texttt{k = 1.0\\ Rc = 5.0\\ gamma = 0.01\\ Rt = 50.0\\ mu = 0.5\\ eta
= 3.0\\ b = 0.0001\\\\ }  If  surface density
data exists in a file named, e.g., \texttt{data.txt}, with columns $R$,
$\Sigma(R)$ and $\Delta\Sigma(R)$, the final fitting is carried out  by use of
the command\\\\ \texttt{ fit f(x) $\prime$data.txt$\prime$
u 1:2:3 via k, Rc, gamma, Rt, eta, mu, b}.

\section{Data tables}
\begin{table*}
\begin{minipage}{168mm}
\centering
 \caption{Fitting results for the edge radii in units of the
 theoretical Jacobi radius for the clusters in a constant tidal field
 averaged over all available snapshots of the specific model. The
 results are given for the three smallest fit ranges. In the last line
 a weighted mean and the standard deviation of the results is
 given. $R_t$ gives the edge radius of the specific template, whereas
 $R_J$ stands for the theoretical Jacobi radius. $\mu$ is the fraction
 of the edge radius, at which the KKBH template turns into a
 power-law.}
\label{table3}
\begin{tabular}{cccccccc}
\hline Name & fit range & King & Gen. Nuker & \multicolumn{2}{c}{KKBH}
& King (bound) \\ & $[ \mbox{pc}]$ & $\frac{R_t}{R_J}$ &
$\frac{R_t}{R_J}$ & $\frac{\mu R_t}{R_J}$& $\frac{ R_t}{R_J}$&
$\frac{R_t}{R_J}$\\ \hline A0 & 1-25 & 0.86 $\pm$ 0.06 & 3.70 $\pm$
0.86 & 0.49 $\pm$ 0.09 & 0.82 $\pm$ 0.07 & 0.77 $\pm$ 0.47\\ & 1-50 &
0.97 $\pm$ 0.06 & 2.12 $\pm$ 0.98 & 0.53 $\pm$ 0.08 & 0.85 $\pm$ 0.07
& 0.74 $\pm$ 0.13\\ & 1-100 & 0.98 $\pm$ 0.06 & 2.22 $\pm$ 0.97 & 0.57
$\pm$ 0.11 & 0.86 $\pm$ 0.08 & 0.74 $\pm$ 0.13\\ B0 & 1-25 & 0.68
$\pm$ 0.05 & 2.58 $\pm$ 0.41 & 0.41 $\pm$ 0.12 & 0.75 $\pm$ 0.06 &
0.70 $\pm$ 0.04\\ & 1-50 & 0.89 $\pm$ 0.05 & 1.48 $\pm$ 0.70 & 0.49
$\pm$ 0.06 & 0.81 $\pm$ 0.06 & 0.71 $\pm$ 0.04\\ & 1-100 & 0.92 $\pm$
0.07 & 1.36 $\pm$ 0.49 & 0.51 $\pm$ 0.07 & 0.81 $\pm$ 0.06 & 0.71
$\pm$ 0.04\\ C0 & 1-25 & 0.63 $\pm$ 0.04 & 1.94 $\pm$ 0.35 & 0.60
$\pm$ 0.05 & 0.76 $\pm$ 0.03 & 0.67 $\pm$ 0.02\\ & 1-50 & 0.84 $\pm$
0.01 & 1.56 $\pm$ 0.89 & 0.46 $\pm$ 0.05 & 0.79 $\pm$ 0.03 & 0.71
$\pm$ 0.01\\ & 1-100 & 0.92 $\pm$ 0.05 & 1.36 $\pm$ 0.74 & 0.50 $\pm$
0.05 & 0.81 $\pm$ 0.03 & 0.71 $\pm$ 0.01\\ D0 & 1-25 & 0.56 $\pm$ 0.05
& 1.58 $\pm$ 0.26 & 0.57 $\pm$ 0.02 & 0.71 $\pm$ 0.02 & 0.63 $\pm$
0.01\\ & 1-50 & 0.78 $\pm$ 0.01 & 1.32 $\pm$ 0.96 & 0.39 $\pm$ 0.03 &
0.76 $\pm$ 0.02 & 0.71 $\pm$ 0.01\\ & 1-100 & 0.90 $\pm$ 0.05 & 1.20
$\pm$ 0.74 & 0.49 $\pm$ 0.05 & 0.80 $\pm$ 0.03 & 0.71 $\pm$ 0.03\\
\hline mean & &0.81 $\pm$ 0.14 & 1.84 $\pm$ 0.72 & 0.50 $\pm$ 0.06 &
0.78 $\pm$ 0.15 & 0.69 $\pm$ 0.04\\
\end{tabular}
\end{minipage}
\end{table*}

\begin{table*}
\begin{minipage}{168mm}
\centering
 \caption{Fitting results for the edge radii in units of the
 theoretical Jacobi radius for the clusters in time-dependent tidal
 fields and orbital eccentricities of 0.25 averaged over all available
 snapshots of the specific model. The results are given for the three
 smallest fit ranges. In the last line a weighted mean and the
 standard deviation of the results is given. $R_t$ gives the edge
 radius of the specific template, whereas $R_J$ stands for the
 theoretical Jacobi radius. $\mu$ is the fraction of the edge radius,
 at which the KKBH template turns into a power-law.}
\label{table4a}
\begin{tabular}{ccccccc}
\hline Name & fit range & King & Gen. Nuker & \multicolumn{2}{c}{KKBH}
& King (bound) \\ & $[ \mbox{pc}]$ & $\frac{R_t}{R_J}$ &
$\frac{R_t}{R_J}$ & $\frac{\mu R_t}{R_J}$& $\frac{ R_t}{R_J}$&
$\frac{R_t}{R_J}$\\ \hline A1 & 1-25 & 0.90 $\pm$ 0.19 & 3.83 $\pm$
1.76 & 0.56 $\pm$ 0.13 & 0.84 $\pm$ 0.17 & 0.71 $\pm$ 0.42\\ & 1-50 &
0.98 $\pm$ 0.22 & 3.37 $\pm$ 1.51 & 0.58 $\pm$ 0.13 & 0.86 $\pm$ 0.17
& 0.69 $\pm$ 0.04\\ & 1-100 & 0.98 $\pm$ 0.22 & 2.08 $\pm$ 1.59 & 0.65
$\pm$ 0.14 & 0.88 $\pm$ 0.18 & 0.69 $\pm$ 0.04\\ B1 & 1-25 & 0.76
$\pm$ 0.12 & 2.97 $\pm$ 0.70 & 0.44 $\pm$ 0.11 & 0.80 $\pm$ 0.13 &
0.71 $\pm$ 0.04\\ & 1-50 & 0.95 $\pm$ 0.16 & 1.60 $\pm$ 0.69 & 0.54
$\pm$ 0.10 & 0.86 $\pm$ 0.14 & 0.72 $\pm$ 0.08\\ & 1-100 & 0.97 $\pm$
0.17 & 1.54 $\pm$ 0.57 & 0.55 $\pm$ 0.12 & 0.85 $\pm$ 0.15 & 0.72
$\pm$ 0.08\\ C1 & 1-25 & 0.66 $\pm$ 0.06 & 2.18 $\pm$ 0.41 & 0.63
$\pm$ 0.09 & 0.81 $\pm$ 0.08 & 0.68 $\pm$ 0.03\\ & 1-50 & 0.90 $\pm$
0.09 & 1.59 $\pm$ 0.94 & 0.51 $\pm$ 0.07 & 0.85 $\pm$ 0.08 & 0.72
$\pm$ 0.02\\ & 1-100 & 0.96 $\pm$ 0.11 & 1.38 $\pm$ 0.53 & 0.54 $\pm$
0.07 & 0.86 $\pm$ 0.08 & 0.72 $\pm$ 0.02\\ D1 & 1-25 & 0.69 $\pm$ 0.13
& 1.86 $\pm$ 0.49 & 0.63 $\pm$ 0.11 & 0.79 $\pm$ 0.13 & 0.66 $\pm$
0.05\\ & 1-50 & 0.87 $\pm$ 0.15 & 1.49 $\pm$ 1.00 & 0.47 $\pm$ 0.09 &
0.82 $\pm$ 0.14 & 0.71 $\pm$ 0.03\\ & 1-100 & 0.97 $\pm$ 0.18 & 1.30
$\pm$ 0.53 & 0.53 $\pm$ 0.09 & 0.85 $\pm$ 0.14 & 0.71 $\pm$ 0.03\\
\hline mean & &0.85 $\pm$ 0.12 & 1.92 $\pm$ 0.84 & 0.55 $\pm$ 0.06 &
0.84 $\pm$ 0.03 & 0.70 $\pm$ 0.02\\
\end{tabular}
\end{minipage}
\end{table*}

\begin{table*}
\begin{minipage}{168mm}
\centering
 \caption{Fitting results for the edge radii in units of the
 theoretical Jacobi radius for the clusters in time-dependent tidal
 fields and orbital eccentricities of 0.5 averaged over all available
 snapshots of the specific model. The results are given for the three
 smallest fit ranges. In the last line a weighted mean and the
 standard deviation of the results is given. $R_t$ gives the edge
 radius of the specific template, whereas $R_J$ stands for the
 theoretical Jacobi radius. $\mu$ is the fraction of the edge radius,
 at which the KKBH template turns into a power-law.}
\label{table4b}
\begin{tabular}{ccccccc}
\hline Name & fit range & King & Gen. Nuker & \multicolumn{2}{c}{KKBH}
& King (bound) \\ & $[ \mbox{pc}]$ & $\frac{R_t}{R_J}$ &
$\frac{R_t}{R_J}$ & $\frac{\mu R_t}{R_J}$& $\frac{ R_t}{R_J}$&
$\frac{R_t}{R_J}$\\ \hline A2 & 1-25 & 1.01 $\pm$ 0.53 & 3.39 $\pm$
2.75 & 0.54 $\pm$ 0.25 & 0.83 $\pm$ 0.40 & 0.61 $\pm$ 0.10\\ & 1-50 &
1.18 $\pm$ 0.68 & 2.75 $\pm$ 2.45 & 0.58 $\pm$ 0.27 & 0.95 $\pm$ 0.68
& 0.62 $\pm$ 0.10\\ & 1-100 & 1.23 $\pm$ 0.73 & 3.46 $\pm$ 2.49 & 0.68
$\pm$ 0.56 & 1.08 $\pm$ 1.04 & 0.62 $\pm$ 0.10\\ B2 & 1-25 & 0.80
$\pm$ 0.31 & 2.72 $\pm$ 1.72 & 0.50 $\pm$ 0.19 & 0.79 $\pm$ 0.30 &
0.66 $\pm$ 0.07\\ & 1-50 & 0.94 $\pm$ 0.37 & 2.30 $\pm$ 1.63 & 0.51
$\pm$ 0.18 & 0.81 $\pm$ 0.31 & 0.66 $\pm$ 0.07\\ & 1-100 & 0.97 $\pm$
0.38 & 1.58 $\pm$ 1.29 & 0.55 $\pm$ 0.19 & 0.82 $\pm$ 0.31 & 0.66
$\pm$ 0.07\\ C2 & 1-25 & 0.70 $\pm$ 0.24 & 2.59 $\pm$ 1.18 & 0.45
$\pm$ 0.17 & 0.76 $\pm$ 0.25 & 0.65 $\pm$ 0.11\\ & 1-50 & 0.90 $\pm$
0.32 & 1.92 $\pm$ 0.98 & 0.51 $\pm$ 0.17 & 0.82 $\pm$ 0.28 & 0.67
$\pm$ 0.07\\ & 1-100 & 0.95 $\pm$ 0.34 & 1.59 $\pm$ 1.02 & 0.54 $\pm$
0.18 & 0.82 $\pm$ 0.28 & 0.67 $\pm$ 0.07\\ D2 & 1-25 & 0.64 $\pm$ 0.22
& 2.25 $\pm$ 0.88 & 0.52 $\pm$ 0.19 & 0.74 $\pm$ 0.25 & 0.62 $\pm$
0.10\\ & 1-50 & 0.87 $\pm$ 0.31 & 1.57 $\pm$ 0.91 & 0.49 $\pm$ 0.16 &
0.79 $\pm$ 0.27 & 0.66 $\pm$ 0.07\\ & 1-100 & 0.94 $\pm$ 0.35 & 1.33
$\pm$ 0.74 & 0.55 $\pm$ 0.18 & 0.81 $\pm$ 0.27 & 0.66 $\pm$ 0.07\\
\hline mean & &0.88 $\pm$ 0.17 & 2.06 $\pm$ 0.72 & 0.52 $\pm$ 0.06 &
0.81 $\pm$ 0.09 & 0.65 $\pm$ 0.02\\
\end{tabular}
\end{minipage}
\end{table*}

\begin{table*}
\begin{minipage}{168mm}
\centering
 \caption{Fitting results for the edge radii in units of the
 theoretical Jacobi radius for the clusters in time-dependent tidal
 fields and orbital eccentricities of 0.75 averaged over all available
 snapshots of the specific model. The results are given for the three
 smallest fit ranges. In the last line a weighted mean and the
 standard deviation of the results is given. $R_t$ gives the edge
 radius of the specific template, whereas $R_J$ stands for the
 theoretical Jacobi radius. $\mu$ is the fraction of the edge radius,
 at which the KKBH template turns into a power-law.}
\label{table4c}
\begin{tabular}{ccccccc}
\hline Name & fit range & King & Gen. Nuker & \multicolumn{2}{c}{KKBH}
& King (bound) \\ & $[ \mbox{pc}]$ & $\frac{R_t}{R_J}$ &
$\frac{R_t}{R_J}$ & $\frac{\mu R_t}{R_J}$& $\frac{ R_t}{R_J}$&
$\frac{R_t}{R_J}$\\ \hline A3 & 1-25 & 2.06 $\pm$ 2.01 & 4.99 $\pm$
2.44 & 1.16 $\pm$ 2.23 & 1.08 $\pm$ 0.72 & 1.47 $\pm$ 2.06\\ & 1-50 &
2.26 $\pm$ 2.27 & 3.67 $\pm$ 2.31 & 1.20 $\pm$ 1.47 & 2.07 $\pm$ 2.44
& 1.43 $\pm$ 1.90\\ & 1-100 & 2.17 $\pm$ 2.09 & 4.12 $\pm$ 2.05 & 1.91
$\pm$ 2.23 & 1.31 $\pm$ 1.28 & 1.43 $\pm$ 1.90\\ B3 & 1-25 & 0.64
$\pm$ 0.22 & 2.25 $\pm$ 0.88 & 0.52 $\pm$ 0.19 & 0.74 $\pm$ 0.25 &
0.62 $\pm$ 0.10\\ & 1-50 & 0.87 $\pm$ 0.31 & 1.57 $\pm$ 0.91 & 0.49
$\pm$ 0.16 & 0.79 $\pm$ 0.27 & 0.66 $\pm$ 0.07\\ & 1-100 & 0.94 $\pm$
0.35 & 1.33 $\pm$ 0.74 & 0.55 $\pm$ 0.18 & 0.81 $\pm$ 0.27 & 0.66
$\pm$ 0.07\\ C3 & 1-25 & 1.38 $\pm$ 1.39 & 3.12 $\pm$ 2.11 & 0.26
$\pm$ 1.68 & 1.60 $\pm$ 1.78 & 0.95 $\pm$ 1.21\\ & 1-50 & 1.96 $\pm$
1.81 & 2.57 $\pm$ 1.94 & 0.88 $\pm$ 1.06 & 1.57 $\pm$ 1.53 & 0.85
$\pm$ 0.73\\ & 1-100 & 2.24 $\pm$ 1.83 & 2.60 $\pm$ 1.91 & 1.04 $\pm$
1.40 & 1.77 $\pm$ 1.84 & 0.87 $\pm$ 0.76\\ D3 & 1-25 & 1.36 $\pm$ 1.67
& 2.67 $\pm$ 1.85 & 1.17 $\pm$ 1.73 & 1.43 $\pm$ 1.53 & 1.07 $\pm$
1.15\\ & 1-50 & 1.75 $\pm$ 1.66 & 2.09 $\pm$ 1.73 & 1.25 $\pm$ 1.94 &
1.74 $\pm$ 1.89 & 1.25 $\pm$ 1.47\\ & 1-100 & 2.06 $\pm$ 1.74 & 2.10
$\pm$ 1.64 & 1.36 $\pm$ 1.80 & 1.74 $\pm$ 1.65 & 1.29 $\pm$ 1.52\\
\hline mean & &1.14 $\pm$ 0.58 & 2.43 $\pm$ 1.06 & 0.67 $\pm$ 0.46 &
1.03 $\pm$ 0.44 & 0.72 $\pm$ 0.32\\
\end{tabular}
\end{minipage}
\end{table*}

\begin{table*}
\begin{minipage}{168mm}
\centering
 \caption{Fitting results for the edge radii in units of the
 theoretical Jacobi radius for the concentrated clusters with a
 half-mass radius of 3 pc averaged over all available snapshots of the
 specific model. The results are given for the three smallest fit
 ranges. In the last line a weighted mean and the standard deviation
 of the results is given. $R_t$ gives the edge radius of the specific
 template, whereas $R_J$ stands for the theoretical Jacobi
 radius. $\mu$ is the fraction of the edge radius, at which the KKBH
 template turns into a power-law.}
\label{table5}
\begin{tabular}{ccccccc}
\hline Name & fit range & King & Gen. Nuker & \multicolumn{2}{c}{KKBH}
& King (bound) \\ & $[ \mbox{pc}]$ & $\frac{R_t}{R_J}$ &
$\frac{R_t}{R_J}$ & $\frac{\mu R_t}{R_J}$& $\frac{ R_t}{R_J}$&
$\frac{R_t}{R_J}$\\ \hline A0c & 1-25 & 0.75 $\pm$ 0.10 & 3.72 $\pm$
0.24 & 0.35 $\pm$ 0.10 & 0.65 $\pm$ 0.11 & 0.79 $\pm$ 0.92\\ & 1-50 &
0.79 $\pm$ 0.10 & 1.85 $\pm$ 0.93 & 0.43 $\pm$ 0.10 & 0.72 $\pm$ 0.11
& 0.69 $\pm$ 0.06\\ & 1-100 & 0.80 $\pm$ 0.10 & 2.46 $\pm$ 0.90 & 0.44
$\pm$ 0.12 & 0.71 $\pm$ 0.12 & 0.69 $\pm$ 0.06\\ B0c & 1-25 & 0.56
$\pm$ 0.10 & 2.51 $\pm$ 0.25 & 0.36 $\pm$ 0.23 & 0.55 $\pm$ 0.19 &
0.59 $\pm$ 0.10\\ & 1-50 & 0.64 $\pm$ 0.12 & 2.17 $\pm$ 0.90 & 0.33
$\pm$ 0.13 & 0.57 $\pm$ 0.14 & 0.64 $\pm$ 0.43\\ & 1-100 & 0.64 $\pm$
0.13 & 1.50 $\pm$ 0.51 & 0.33 $\pm$ 0.09 & 0.58 $\pm$ 0.14 & 0.64
$\pm$ 0.43\\ C0c & 1-25 & 0.47 $\pm$ 0.12 & 1.87 $\pm$ 0.60 & 0.34
$\pm$ 0.19 & 0.49 $\pm$ 0.19 & 0.52 $\pm$ 0.12\\ & 1-50 & 0.55 $\pm$
0.14 & 1.48 $\pm$ 0.57 & 0.20 $\pm$ 0.06 & 0.43 $\pm$ 0.12 & 0.53
$\pm$ 0.13\\ & 1-100 & 0.56 $\pm$ 0.15 & 1.39 $\pm$ 0.34 & 0.26 $\pm$
0.10 & 0.49 $\pm$ 0.15 & 0.53 $\pm$ 0.13\\ D0c & 1-25 & 0.36 $\pm$
0.06 & 1.65 $\pm$ 1.04 & 0.26 $\pm$ 0.15 & 0.38 $\pm$ 0.14 & 0.41
$\pm$ 0.07\\ & 1-50 & 0.42 $\pm$ 0.09 & 1.36 $\pm$ 0.81 & 0.14 $\pm$
0.02 & 0.32 $\pm$ 0.06 & 0.42 $\pm$ 0.09\\ & 1-100 & 0.43 $\pm$ 0.10 &
1.29 $\pm$ 0.32 & 0.17 $\pm$ 0.05 & 0.36 $\pm$ 0.08 & 0.42 $\pm$
0.09\\ \hline mean & &0.57 $\pm$ 0.15 & 2.10 $\pm$ 0.70 & 0.24 $\pm$
0.10 & 0.50 $\pm$ 0.13 & 0.55 $\pm$ 0.12\\
\end{tabular}
\end{minipage}
\end{table*}

\bsp \label{lastpage} 
\begin{thebibliography}{99}

\bibitem[\protect\citeauthoryear{Aarseth}{2003}]{Aarseth03} Aarseth
S.~J., 2003, Gravitational N-Body Simulations, Cambridge, UK,
Cambridge University Press

\bibitem[\protect\citeauthoryear{Allen \& Santillan}{1991}]{Allen91}
Allen C., Santillan C., 1991, RMxAA, 22, 255

\bibitem[\protect\citeauthoryear{Barmby et al.}{2009}]{Barmby09}
Barmby P., et al., 2009, AJ, 138, 1667

\bibitem[\protect\citeauthoryear{Baumgardt}{2001}]{Baumgardt01}
Baumgardt H., 2001, MNRAS, 325, 1323

\bibitem[\protect\citeauthoryear{Baumgardt \&
Makino}{2003}]{Baumgardt03} Baumgardt H., Makino J., 2003, MNRAS, 340,
227

\bibitem[\protect\citeauthoryear{Baumgardt et al.}{2009}]{Baumgardt09}
Baumgardt H., C{\^o}t{\'e} P., Hilker M., Rejkuba M., Mieske S.,
Djorgovski S.~G., Stetson P., 2009, MNRAS, 396, 2051

\bibitem[\protect\citeauthoryear{Baumgardt et al.}{2010}]{Baumgardt10}
Baumgardt H., Parmentier G., Gieles M., Vesperini E., 2010, MNRAS,
401, 1832

\bibitem[\protect\citeauthoryear{Binney \& Tremaine}{2008}]{Binney87}
Binney J., Tremaine S., 2008, Galactic Dynamics, Princeton, NJ,
Princeton University Press
 
\bibitem[\protect\citeauthoryear{Bonatto \& Bica}{2008}]{Bonatto08}
Bonatto C., Bica E., 2008, A\&A, 477, 829

\bibitem[\protect\citeauthoryear{Capuzzo Dolcetta, Di Matteo, \&
Miocchi}{2005}]{Capuzzo05} Capuzzo Dolcetta R., Di Matteo P., Miocchi
P., 2005, AJ, 129, 1906

\bibitem[\protect\citeauthoryear{Carraro, Zinn \& Moni
Bidin}{2007}]{Carraro07} Carraro G., Zinn R., Moni Bidin C., 2007,
A\&A, 466, 181

\bibitem[\protect\citeauthoryear{Carraro}{2009}]{Carraro09} Carraro
G., 2009, AJ, 137, 3809

\bibitem[\protect\citeauthoryear{Casertano \& Hut}{1985}]{Casertano85}
Casertano S., Hut P., 1985, ApJ, 298, 80

\bibitem[\protect\citeauthoryear{C{\^o}t{\'e} et al.}{2002}]{Cote02}
C{\^o}t{\'e} P., Djorgovski S.~G., Meylan G., Castro S., McCarthy
J.~K., 2002, ApJ, 574, 783

\bibitem[\protect\citeauthoryear{Dehnen et al.}{2004}]{Dehnen04}
Dehnen W., Odenkirchen M., Grebel E.~K., Rix H.-W., 2004, AJ, 127,
2753

\bibitem[\protect\citeauthoryear{Drukier et al.}{1998}]{Drukier98}
Drukier G.~A., Slavin S.~D., Cohn H.~N., Lugger P.~M., Berrington
R.~C., Murphy B.~W., Seitzer P.~O., 1998, AJ, 115, 708

\bibitem[\protect\citeauthoryear{Drukier et al.}{2007}]{Drukier07}
Drukier G.~A., Cohn H.~N., Lugger P.~M., Slavin S.~D., Berrington
R.~C., Murphy B.~W., 2007, AJ, 133, 1041

\bibitem[\protect\citeauthoryear{Elson, Fall \&
Freeman}{1987}]{Elson87} Elson R.~A.~W., Fall S.~M., Freeman K.~C.,
1987, ApJ, 323, 54

\bibitem[\protect\citeauthoryear{Fall \& Zhang}{2001}]{Fall01} Fall
S.~M., Zhang Q., 2001, ApJ, 561, 751

\bibitem[\protect\citeauthoryear{Fukushige \&
Heggie}{2000}]{Fukushige00} Fukushige T., Heggie D.~C., 2000, MNRAS,
318, 753

\bibitem[\protect\citeauthoryear{Fukushige, Makino \&
Kawai}{2005}]{Fukushige05} Fukushige T., Makino J., Kawai A., 2005,
PASJ, 57, 1009

\bibitem[\protect\citeauthoryear{Gieles, Sana \& Portegies
Zwart}{2009}]{Gieles09} Gieles M., Sana H., Portegies~Zwart S.~F.,
2009, MNRAS, in press

\bibitem[\protect\citeauthoryear{Gnedin, Lee \&
Ostriker}{1999}]{Gnedin99} Gnedin O.~Y., Lee H.~M., Ostriker J.~P.,
1999, ApJ, 522, 935

\bibitem[\protect\citeauthoryear{Gouliermis et
al. }{2010}]{Gouliermis10} Gouliermis D.~A., Mackey D., Xin Y., Rochau
B., 2010, ApJ, 709, 263

\bibitem[\protect\citeauthoryear{Heggie \& Hut}{2003}]{Heggie03}
Heggie D., Hut P., 2003, The Gravitational Million-Body Problem,
Cambridge, UK, Cambridge University Press

\bibitem[\protect\citeauthoryear{Innanen, Harris, \&
Webbink}{1983}]{Innanen83} Innanen K.~A., Harris W.~E., Webbink R.~F.,
1983, AJ, 88, 338

\bibitem[\protect\citeauthoryear{Jordi et al.}{2009}]{Jordi09} Jordi
K., et al., 2009, AJ, 137, 4586

\bibitem[\protect\citeauthoryear{Just et al.}{2009}]{Just09} Just A.,
Berczik P., Petrov M.~I., Ernst A., 2009, MNRAS, 392, 969

\bibitem[\protect\citeauthoryear{King}{1962}]{King62} King I.~R.,
1962, AJ, 67, 471

\bibitem[\protect\citeauthoryear{King}{1966}]{King66} King I.~R.,
1966, AJ, 71, 64

\bibitem[\protect\citeauthoryear{Kouwenhoven \& de
Grijs}{2009}]{Kouwenhoven09} Kouwenhoven M.~B.~N., de Grijs R., 2009,
Ap\&SS, 324, 171

\bibitem[\protect\citeauthoryear{Kroupa}{2001}]{Kroupa01} Kroupa P.,
2001, MNRAS, 322, 231

\bibitem[\protect\citeauthoryear{K\"upper, Macleod \&
Heggie}{2008}]{Kuepper08a} K\"upper A.~H.~W., Macleod A., Heggie
D.~C., 2008, MNRAS, 387, 1248

\bibitem[\protect\citeauthoryear{K\"upper, Kroupa \&
Baumgardt}{2008}]{Kuepper08b} K\"upper A.~H.~W., Kroupa P., Baumgardt
H., 2008, MNRAS, 389, 889

\bibitem[\protect\citeauthoryear{K\"upper et al.}{2010}]{Kuepper09}
K\"upper A.~H.~W., Kroupa P., Baumgardt H., Heggie D.~C., 2010, MNRAS,
401, 105

\bibitem[\protect\citeauthoryear{K\"upper \&
Kroupa}{2010}]{Kuepper10b} K\"upper A.~H.~W., Kroupa P., 2010, ApJ
submitted

\bibitem[\protect\citeauthoryear{Lane et al.}{2009}]{Lane09} Lane
R.~R., Kiss L.~L., Lewis G.~F., Ibata R.~A., Siebert A., Bedding
T.~R., Sz{\'e}kely P., 2009, MNRAS, 400, 917

\bibitem[\protect\citeauthoryear{Lane et al.}{2010}]{Lane10} Lane
R.~R., Kiss L.~L., Lewis G.~F., Ibata R.~A., Siebert A., Bedding
T.~R., Sz{\'e}kely P., 2010, MNRAS, 401, 2521

\bibitem[\protect\citeauthoryear{Lauer et al.}{1995}]{Lauer95} Lauer
T.~R., et al., 1995, AJ, 110, 2622

\bibitem[\protect\citeauthoryear{McLaughlin}{2003a}]{McLaughlin03a}
McLaughlin D.~E., 2003, ASPC, 296, 101

\bibitem[\protect\citeauthoryear{McLaughlin \&
Meylan}{2003b}]{McLaughlin03b} McLaughlin D.~E., Meylan G., 2003,
ASPC, 296, 153

\bibitem[\protect\citeauthoryear{McLaughlin \& van der
Marel}{2005}]{McLaughlin05} McLaughlin D.~E., van der Marel R.~P.,
2005, ApJS, 161, 304

\bibitem[\protect\citeauthoryear{Milgrom}{1983}]{Milgrom83} Milgrom
M., 1983, ApJ, 270, 365

\bibitem[\protect\citeauthoryear{Odenkirchen et
al.}{2003}]{Odenkirchen03} Odenkirchen M., et al., 2003, AJ, 126, 2385

\bibitem[\protect\citeauthoryear{Scarpa et al.}{2007}]{Scarpa07}
Scarpa R., Marconi G., Gilmozzi R., Carraro G., 2007, Msngr, 128, 41

\bibitem[\protect\citeauthoryear{Scarpa, Marconi \&
Gilmozzi}{2003}]{Scarpa03} Scarpa R., Marconi G., Gilmozzi R., 2003,
A\&A, 405, L15

\bibitem[\protect\citeauthoryear{Spitzer}{1987}]{Spitzer87} Spitzer
L., 1987, Dynamical evolution of globular clusters, Princeton, NJ,
Princeton University Press, 1987, 191 p.

\bibitem[\protect\citeauthoryear{{\v S}ubr, Kroupa, \&
Baumgardt}{2008}]{Subr08} {\v S}ubr L., Kroupa P., Baumgardt H., 2008,
MNRAS, 385, 1673

\bibitem[\protect\citeauthoryear{Trenti, Vesperini \&
Pasquato}{2010}]{Trenti10} Trenti M., Vesperini E., Pasquato M., 2010,
ApJ, 708, 1598

\bibitem[\protect\citeauthoryear{van der Marel \&
Anderson}{2010}]{vanderMarel09} van der Marel R.~P., Anderson J.,
2010, ApJ, 710, 1063

\bibitem[\protect\citeauthoryear{Wilson}{1975}]{Wilson75} Wilson
C.~P., 1975, AJ, 80, 175


\end{thebibliography}
\end{document}